\documentclass[aps,floatfix]{revtex4}
\usepackage{graphics}
\usepackage{color}
\usepackage{graphicx}% Include figure files
\usepackage{bm}
\begin{document}
\title{Cubic-quintic nonlinearity in superfluid Bose-Bose mixtures in optical lattices: Heavy solitary waves, barrier-induced criticality, and current-phase relations}
\author{Ippei Danshita$^{1,2}$}
\author{Daisuke Yamamoto$^3$}
\author{Yasuyuki Kato$^4$}
\affiliation{$^1$Yukawa Institute for Theoretical Physics, Kyoto University, Kyoto 606-8502, Japan \\
$^2$Computational Condensed Matter Physics Laboratory, RIKEN, Wako, Saitama 351-0198, Japan \\
$^3$Waseda Institute for Advanced Study, Waseda University, Tokyo 169-8050, Japan \\
$^4$RIKEN Center for Emergent Matter Science (CEMS), Wako, Saitama 351-0198, Japan}
\date{\today}

\begin{abstract}
We study superfluid (SF) states of strongly interacting Bose-Bose mixtures with equal mass and intra-component interaction in optical lattices both in the presence and absence of a barrier potential. We show that the SF order parameters obey the two-component nonlinear Schr\"odinger equation (NLSE) with not only cubic but also quintic nonlinearity in the vicinity of the first-order transitions to the Mott insulators with even fillings. 
In the case of no barrier potential, we analyze solitary-wave solutions of the cubic-quintic NLSE. When the SF state changes from a ground state to a metastable one, a standard dark solitary wave turns into a bubble-like dark solitary wave, which has a non-vanishing density dip and no $\pi$ phase kink even in the case of a standing solitary wave. It is shown that the former and latter solitary waves are dynamically unstable against an out-of-phase fluctuation and an in-phase fluctuation, respectively, and the dynamical instabilities are weakened when one approaches the transition point. We find that the size and the inertial mass of the solitary waves diverge at the first-order transition point. We suggest that the divergence of the inertial mass may be detected through measurement of the relation between the velocity and the phase jump of the solitary wave.
In the presence of a barrier potential, we reveal that when the barrier strength exceeds a certain critical value, the SF state that was metastable without the barrier is destabilized towards complete disjunction of the SF. The presence of the critical barrier strength indicates that the strong barrier potential qualitatively changes the criticality near the metastability limit of the SF state. We derive critical behaviors of the density, the compressibility, and the critical current near the metastability limit induced by the barrier. It is also found that the relation between the supercurrent and the phase jump across the barrier exhibits a peculiar behavior, owing to the non-topological nature of the bubble-like solitary wave.

\end{abstract}

\maketitle

\section{Introduction}
Since the realization of Bose-Einstein condensates (BEC) of alkali atomic gases~\cite{anderson-95,davis-95}, experiments studying BEC have been extensively compared with microscopic theories, and it has been established that BEC of a weakly-interacting dilute Bose gas near zero temperature can be quantitatively described by the Gross-Pitaevskii (GP) equation that is a type of nonlinear Schr\"odinger equation (NLSE) with cubic nonlinearity~\cite{dalfovo-99, pitaevskii-03}. The wide applicability of the GP equation allows one to predict BEC properties not only for ground states but also for excited states and non-equilibrium dynamics. Despite its simple form, the cubic nonlinearity originated from the contact interparticle interaction gives rise to a large number of intriguing effects and phenomena regarding BEC, such as the Bogoliubov excitation spectrum~\cite{steinhauer-02}, nonlinear couplings between different collective modes~\cite{hechenblaikner-00}, bright~\cite{khaykovich-02} and dark solitons~\cite{burger-99, denschlag-00, becker-08}, vortices~\cite{madison-01,abo-02}, supercurrent and its breakdown above the critical velocity~\cite{onofrio-00, sarlo-05, engels-07, levy-07, ramanathan-11, desbuquois-12},  and the self-trapped motion in a double-well potential~\cite{albiez-05}.

When a BEC is loaded onto an optical lattice, the interparticle interaction relative to the kinetic energy can be widely controlled so that one can achieve a strongly interacting regime where the superfluid (SF) state does not obey the GP equation any longer. A clear demonstration of this fact is the observation of the quantum phase transition between the SF and the Mott insulator (MI) at commensurate fillings~\cite{greiner-02}, which the GP equation completely fails to capture. Instead, the SF state near the SF-MI transition is described by a Lorentz-invariant version of fourth-order Ginzburg-Landau (GL) theory, whose saddle-point approximation corresponds to a nonlinear Klein-Gordon equation with cubic nonlinearity~\cite{fisher-89, altman-02}. Thanks to the Lorentz invariance, there emerges new properties that are absent in the GP equation, such as Higgs amplitude modes~\cite{altman-02, endres-12, pekker-14} and criticality of the SF critical velocity for dynamical instability~\cite{altman-05, mun-07}. When the interaction is further stronger at low density, the system reaches the hardcore-boson regime~\cite{paredes-04}, where the SF state obeys a Landau-Lifshitz equation that is qualitatively different from the GP equation~\cite{balakrishnan-09}. Thus, strong correlations in the optical-lattice systems may be utilized to design several types of SF that obeys equations of motion other than the GP equation.

In this paper, we show that a SF state of Bose-Bose mixtures in optical lattices obeys a NLSE with cubic-quintic nonlinearity in certain parameter regions. In the previous work by the authors~\cite{kato-14}, the sixth-order GL action has been derived from the two-component Bose-Hubbard model (BHM) in the vicinity of the first-order SF-MI transitions. We apply a saddle-point approximation to the GL action in order to derive the two-component cubic-quintic NLSE. While cubic-quintic NLSE has been analyzed in previous studies in the contexts of first-order phase transitions of condensed-matter systems~\cite{ginzburg-82, lipowsky-82, lipowsky-83, sornette-85, binder-87, barashenkov-88, barashenkov-89} and nonlinear optics~\cite{gagnon-89, kivshar-98, malomed-05}, we emphasize the following three advantages of our optical-lattice system. First, thanks to its exquisite controllability and cleanness, the parameters in the original BHM can be widely varied, e.g., by controlling the lattice depth, the density of the gas, the trapping potential, and other external fields. Second, the parameters in the GL action are controllable as well, because they are explicitly related to those in the original BHM~\cite{kato-14}. Third, the long relaxation time specific to cold-atom systems enables one to study non-equilibrium dynamics in greater details. Having these advantages in mind, we specifically investigate dark solitary waves and barrier-potential effects in the SF state described by the cubic-quintic NLSE. Notice that the advantages mentioned above are relevant also to the recently proposed cold-atom systems with local three-body interactions if a mean-field approximation is applied to the corresponding models~\cite{johnson-09, mahmud-13, petrov-14}.

Existence of solitary waves is one of the simplest but most essential consequences due to nonlinearity. Bright~\cite{khaykovich-02} and dark~\cite{burger-99, denschlag-00, becker-08} solitons of atomic BEC in the GP regime have been observed, and their basic properties have been well understood~\cite{pitaevskii-03, frantzeskakis-10}. To provide clear contrast to the GP solitons, we analyze dark soltary-wave solutions of the two-component cubic-quintic NLSE. Previous studies have shown that there are two types of single dark solitary-wave solution of the cubic-quintic NLSE~\cite{barashenkov-88}. One is a standard dark solitary wave that is a nonlinear excitation of a ground-state SF. It has a $\pi$-phase jump when it is at rest, as in the case of the GP dark soliton. On the other hand, when the SF state is metastable, there emerges a bubble-like solitary wave that has no phase jump at zero velocity. For both types of solitary wave, we analytically calculate the size and the inertial mass to show that they diverge at the first-order SF-MI transition point. On the basis of the direct connection between the inertial mass and the phase jump of dark solitary waves~\cite{scott-11, pitaevskii-14}, we propose a way to observe the divergence of the inertial mass in experiments. 

The divergent behaviors of the solitary waves are remarkable in the sense that they manifest criticality associated with the first-order quantum phase transition, which is not exhibited by linear excitations or thermodynamic quantities of uniform SF states. We also stress the importance of our prediction of such a heavy dark solitary wave in connection with a recent experiment. The experimental group of Zwierlein at MIT has reported the observation of a surprisingly-heavy dark soliton in the system of a SF Fermi gas near the unitarity limit and triggered renewed interest in solitary waves of ultracold gases~\cite{yefsah-13}.   Although it has turned out that the observed object was not a dark soliton but a single vortex line~\cite{ku-14}, the question regarding the possibility of unusually heavy solitary waves still remains. The dark solitary wave with a diverging mass predicted in this paper serves as the first example of such a heavy solitary wave.

When a barrier potential is present in the SF state, we find that a barrier potential stronger than a certain threshold value disrupts the SF state that was metastable without the barrier. The presence of the critical barrier strength leads to the emergence of new criticality at the metastability limit of the SF state, which is often referred to as surface critical phenomena~\cite{lipowsky-82, lipowsky-83, lipowsky-84, sornette-85, lipowsky-87}. This criticality is equivalent to that of the dark solitary waves with no barrier potential in the sense that it accompanies the divergence of the size of the density dip. We point out that although thermodynamic quantities, such as the average density and the compressibility, exhibit the critical behaviors near the barrier-induced metastability limit, the signals are too weak to use for identifying the criticality numerically or experimentally. 

Another important effect on BEC appearing as a consequence of nonlinearity is a supercurrent past a barrier potential, which means that a BEC acquires the superfluidity thanks to the nonlinearity. Previous studies on the GP equation have derived the relation between the supercurrent and the phase jump across the barrier potential and shown that it becomes the cerebrated Josephson relation~\cite{josephson-62} in the strong-barrier regime~\cite{baratoff-70, danshita-06, watanabe-09}. In the case of the cubic-quintic NLSE, we show that the current-phase relation also approaches to the Josephson relation with increasing the barrier strength as long as the current-free SF state in the absence of the barrier is a ground state. In contrast, when the SF state is metastable, the Josephson relation is not held any longer because of the disappearance of the metastable state above the critical barrier strength. Moreover, we calculate the critical current above which a current-carrying state is unstable, in order to show that its critical behavior near the barrier-induced metastability limit gives a stronger signal than the thermodynamic quantities and that it may be useful for experimental detection of the criticality.  

The remainder of the paper is organized as follows. In Sec.~\ref{sec:GL}, we derive the two-component cubic-quintic NLSE from the sixth-order GL action. Parameter regions in which the GL action is valid are presented. In Sec.~\ref{sec:phd}, considering a uniform potential and a uniform solution, we briefly review how the first-order quantum phase transition is described within the sixth-order GL theory. Dark solitary-wave solutions of the cubic-quintic NLSE are analyzed in Sec.~\ref{sec:soliton}, where a special emphasis is placed on the divergence of the size and the inertial mass of the solitary waves. In Sec.~\ref{sec:barrier}, we derive analytical solutions of the cubic-quintic NLSE in the presence of a barrier potential. On the basis of the solutions, we discuss the barrier-induced criticality and the current-phase relation. The results are summarized in Sec.~\ref{sec:conc}

%%%%%%%%%%%%%%%%%%%%%%%
\section{Sixth-order Ginzburg-Landau theory}\label{sec:GL}
%%%%%%%%%%%%%%%%%%%%%%%
We consider a binary Bose mixture confined in a hypercubic optical lattice. We assume a sufficiently deep lattice so that the system is well described by the two-component BHM~\cite{jaksch-98},
\begin{eqnarray}
  \hat{H}=
    \sum_{\alpha} \left[ -\sum_{\bm j}\sum_{\sigma=1}^{d} t_{\alpha}
    \left(
    \hat{b}^{\dagger}_{\alpha, {\bm j}} \hat{b}_{\alpha, {\bm j}+{\bm e}_{\sigma}}
    +{\rm H.c.}
    \right)
    + \sum_{\bm j}\left( \frac{U_{\alpha}}{2}
   \hat{n}_{\alpha,{\bm j}} (\hat{n}_{\alpha,{\bm j}}-1)
    - \mu_{\alpha,{\bm j}} \hat{n}_{\alpha,{\bm j}} \right)
    \right]
+ \sum_{\bm j} U_{AB} \hat{n}_{A,{\bm j}} \hat{n}_{B,{\bm j}}
\label{eq:hamiltonian}
\end{eqnarray}
where ${\bm j}\equiv \sum_{\sigma=1}^{d} j_{\sigma} {\bm e}_{\sigma}$ denotes the site index, $j_{\sigma}$ is an integer, and $d$ is the spatial dimension of the system. ${\bm e}_{\sigma}$ denotes a unit vector in direction $\sigma$, where the directions $\sigma = 1, 2,$ and 3 mean the $x, y$, and $z$ directions.  
$t_{\alpha}$ and $U_{\alpha}$ are the hopping and the intra-component onsite interaction for the component $\alpha\in \{A,B\}$.  The local chemical potential $\mu_{\alpha,{\bm j}}\equiv \mu_\alpha - \epsilon_{\alpha, {\bm j}}$ consists of the global chemical potential $\mu_{\alpha}$ and the external potential $\epsilon_{\alpha,{\bm j}}$. The inter-component interaction is denoted by $U_{AB}$. Hereafter we assume the symmetry with respect to the exchange $A\leftrightarrow B$, i.e., $t_{A} = t_{B}\equiv t$, $U_{A} = U_{B}\equiv U$, $\mu_{A,{\bm j}}=\mu_{B,{\bm j}}\equiv \mu_j$, $\mu_{A}=\mu_{B}\equiv \mu$, and $\epsilon_{A,{\bm j}}=\epsilon_{B,{\bm j}}\equiv \epsilon_{\bm j}$. This condition is nearly satisfied in a mixture of the two hyperfine states $|F=2,m_F=-1\rangle$ and $|F=1,m_F=1\rangle$ of $^{87}$Rb, where the scattering length between two atoms in the former state is only 5\% smaller than that in the latter state~\cite{hall-98, egorov-13}. The $^{87}$Rb mixture is advantageous also in the sense that the inter-component interaction is controllable with use of the Feshbach resonances~\cite{widera-08,tojo-10} or state-dependent optical lattices~\cite{mckay-10}. The ground-state phases of the two-component BHM at $T=0$ are rather rich even in the $A\leftrightarrow B$ symmetric case and have been addressed in previous theoretical studies~\cite{kuklov-03, paredes-03, chen-03, altman-03, kuklov-04-a, kuklov-04-b, isacsson-05, arguelles-07, mishra-07, mathey-09, hu-09, hubener-09, iskin-10, chen-10, sansone-10, ozaki-12, li-13, yamamoto-13, kato-14}. It is well known that the transition from SF to MI occurs when $Zt/U$ decreases or when $\mu/U$ changes for a small $Zt/U$, as in the case of the single-component BHM~\cite{fisher-89}. Here $Z$ is the coordination number. When $d\ge 2$ and $\chi <U_{AB}/U<1$, the transition to MI with even fillings is first order~\cite{kuklov-04-a, ozaki-12}, where the lower bound for $\nu=2$ is evaluated as $\chi \simeq 0.68$ within the Gutzwiller mean-field approximation~\cite{ozaki-12,yamamoto-13}. Here $\nu$ denotes the filling factor.

%%%%%%%%%%%%%%%%%%%%%%%%%%%%%%%%%%%
\begin{figure}[tb]
\includegraphics[scale=0.5]{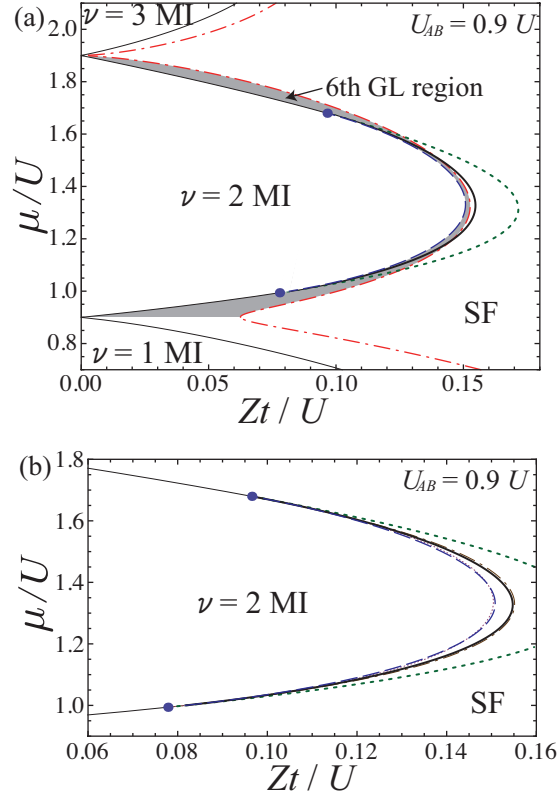}
\caption{\label{fig:sketch}
(color online) Phase diagrams of the two-component BHM of Eq.~(\ref{eq:hamiltonian}) in the $(zt/U, \mu/U)$ plane computed by means of the Gutzwiller mean-field approximation, where $U_{AB}/U=0.9$. The thin-solid and thick-solid lines represent the phase boundaries of the first-order and second-order transitions. The dashed and dotted lines represent the metastability limits of the SF and MI states. The dots mark the TCPs. Those lines and dots are taken from Refs.~\cite{ozaki-12,yamamoto-13}. In (a), the dash-dotted line represents the contour of $|\psi_{A}|^2a^d = |\psi_{B}|^2a^d = 0.25$ and the gray shaded area roughly marks the region where the sixth-order GL theory is validated for describing the SF state. In (b), the thin-dash-dotted and thin-dotted lines represent the first-order phase boundary and the SF metastability limit calculated by the sixth-order GL theory.
}
\end{figure}
%%%%%%%%%%%%%%%%%%%%%%%%%%%%%%%%%%%

In the previous work of the authors~\cite{kato-14}, it has been shown that in the vicinity of the first-order transition points the system is described by the following effective action of the sixth-order GL form, 
\begin{eqnarray}
S^{\rm eff} &=& \int  d\tau \int d^{d}x 
\left[ 
\sum_{\alpha} 
\left(
i\hbar K({\bm x}) \psi_{\alpha}^{\ast} \frac{\partial \psi_{\alpha}}{\partial \tau} 
-\hbar^2 J({\bm x}) \left|\frac{\partial \psi_{\alpha}}{\partial \tau}\right|^2
- \frac{\hbar^2}{2m}|\nabla\psi_{\alpha}|^2   
+r(\bm{x}) |\psi_{\alpha}|^2 
- \frac{u(\bm{x})}{2} |\psi_{\alpha}|^4 
- \frac{w(\bm{x})}{3}|\psi_{\alpha}|^6 \right) \right. \nonumber \\
&& 
\,\,\,\,\,\,\,\,\,\,\,\,\,\,\,\,\,\,\,\,\,\,\,\,\,\,\,\,\,\,\,\,\,\,\,\,\,\,\,\,\,\,
\Biggl.
- u_{AB}(\bm{x}) |\psi_A|^2 |\psi_B|^2
- w_{AB}(\bm{x})\left(|\psi_A|^4 |\psi_B|^2 + |\psi_A|^2 |\psi_B|^4 \right) 
\Biggr],
\label{eq:action}
\end{eqnarray}
where $\psi_{\alpha}(\bm{x},\tau)$ denotes the SF order-parameter field of the component $\alpha$ at $\bm{x} \equiv a{\bm j}$ and real time $\tau$. In Eq.~(\ref{eq:action}), the continuum limit has been taken under the assumption that the lattice spacing $a$ is much smaller than the healing length $\xi$. This effective action well describes the SF state when $|\psi_{\alpha}|^2a^{d} \ll 1$. In the case of a homogeneous system, i.e., $\mu_{\bm j} = \mu$, the parameter region around $\nu= 2$, in which the effective action is approximately valid, is depicted as the gray shaded area in Fig.~\ref{fig:sketch}(a). As seen in Fig.~\ref{fig:sketch}(b), the first-order phase boundary and the SF metastability limit obtained by the GL theory agrees well with those computed by the Gutzwiller mean-field approximation in Refs.~\cite{ozaki-12,yamamoto-13}. If the transition is of second order and sufficiently far from the tricritical point (TCP), at which the first-order transition shifts to the second-order one, the sixth-order terms in the action can be ignored. Otherwise, the ignorance of those terms leads to a qualitative failure of the action.

%%%%%%%%%%%%%%%%%%%%%%%%%%%%%%%%%%%
\begin{figure}[tb]
\includegraphics[scale=0.5]{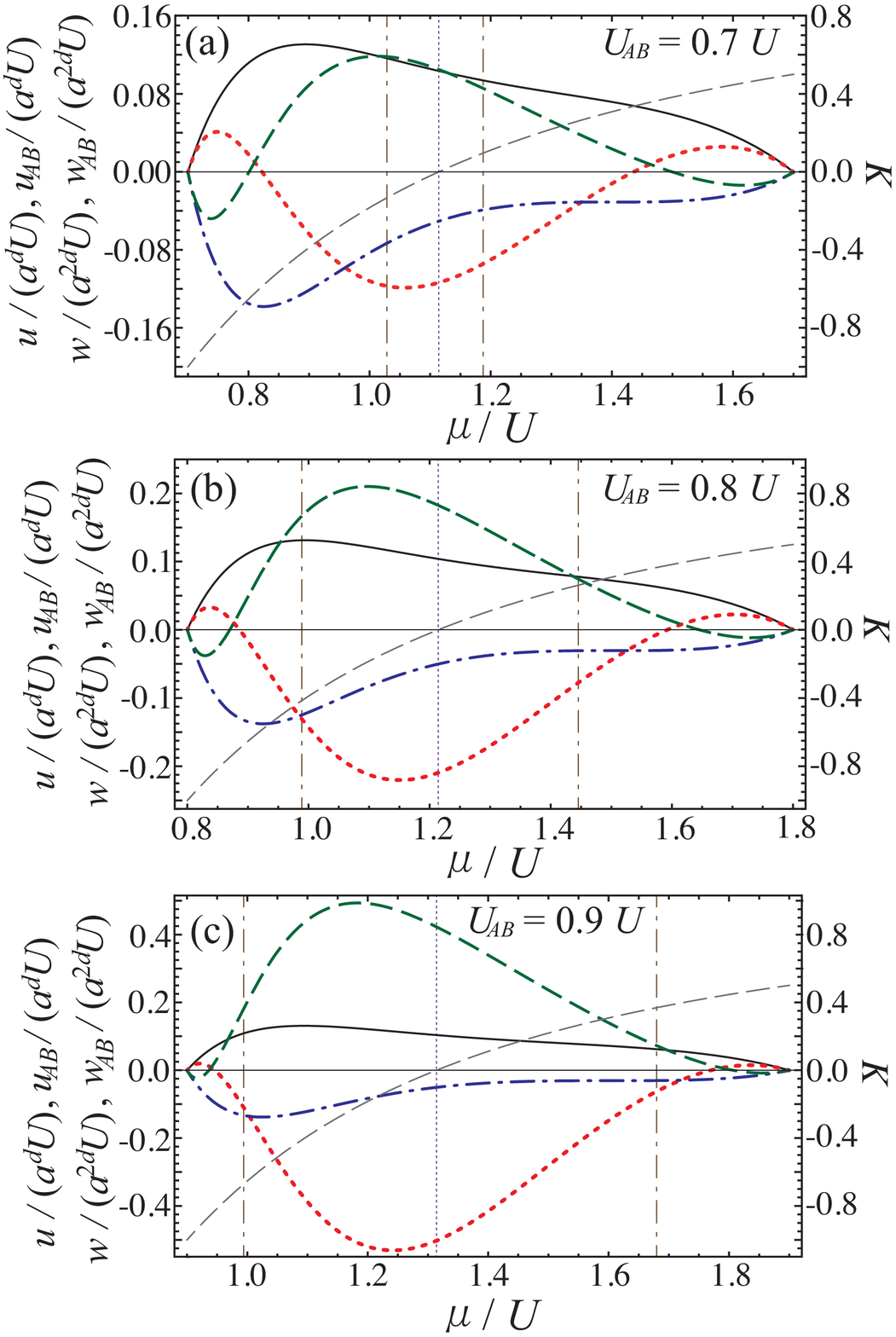}
\caption{\label{fig:uwK}
(color online) GL parameters along the metastability limit of the MI state at $\nu=2$, as functions of $\mu/U$ for $U_{AB} /U = 0.7$ (a), $0.8$ (b), and $0.9$ (c). The solid, thick-dotted, thick-dash-dotted, thick-dashed, and thin-dashed lines represent $u/(a^d U)$, $u_{AB}/(a^d U)$, $w/(a^{2d} U)$, $w_{AB}/(a^{2d} U)$, and $K$, respectively. The thin-dash-dotted and thin-dotted lines mark the TCPs and the $\mu$ value that gives the maximum hopping.
}
\end{figure}
%%%%%%%%%%%%%%%%%%%%%%%%%%%%%%%%%%%

All the coefficients in the effective action are explicitly related to the parameters in the BHM through a perturbative expansion~\cite{kato-14}. More specifically, they are the functions of the filling factor of the MI state $\nu_0$, $Zt/U$, $U_{AB}/U$, and $\mu_{\bm j}/U$.
Obviously, their position dependence stems from that of the local chemical potential.
Assuming that $\delta \mu_{\bm j}/U\ll 1$, one may approximate the coefficients other than $r({\bm x})$ as their values at $\mu_{\bm j} = \mu_{\rm MI}$, namely
$K({\bm x}) \simeq K|_{\mu_{\bm j} = \mu_{\rm MI}}$, $J({\bm x}) \simeq J|_{\mu_{\bm j} = \mu_{\rm MI}}$, $u({\bm x}) \simeq u|_{\mu_{\bm j} = \mu_{\rm MI}}$, $u_{AB}({\bm x}) \simeq u_{AB}|_{\mu_{\bm j} = \mu_{\rm MI}}$, $w({\bm x}) \simeq w|_{\mu_{\bm j} = \mu_{\rm MI}}$, and $w_{AB}({\bm x}) \simeq w_{AB}|_{\mu_{\bm j} = \mu_{\rm MI}}$.
Here $\delta \mu_{\bm j} \equiv \mu_{\bm j} - \mu_{\rm MI}$ and $\mu_{\rm MI}$ denotes the chemical potential value at the metastability limit of the MI state.  
As for the coefficient $r({\bm x})$, since $r|_{\mu_j = \mu_{\rm MI}} = 0$, one needs to include the next-order term as $r({\bm x}) \simeq C \delta \mu_{\bm j}$, where the constant $C$ is positive in the upper side of the Mott lobe while it is negative in the lower side.
We assume that the system is sufficiently far from the tip of the Mott lobe, at which $K=0$, and focus on low-energy physics of the system. In such a situation, the $J$ term can be ignored.
In Fig.~\ref{fig:uwK}, we plot the GL parameters $K$, $u$, $u_{AB}$, $w$, and $w_{AB}$ at $\mu_{\bm j} = \mu_{\rm MI}$ and $\nu=2$ as functions of $\mu_{\rm MI}$ for several values of $U_{AB}/U$.
Hereafter we take the unit of $K=1$, and $u$, $u_{AB}$, $w$, and $w_{AB}$ denote the values at $\mu_{\bm j} = \mu_{\rm MI}$.

Minimizing the action under the condition $\frac{\delta S^{\rm eff}}{\delta \psi_{\alpha} } = 0$ leads to the time-dependent GL equations,
\begin{eqnarray}
i\hbar\frac{\partial \psi_A}{\partial \tau} &=& 
\left[ -\frac{\hbar^2}{2m} \nabla^2 - r({\bm x}) + u|\psi_A|^2 + u_{AB}|\psi_B|^2 
+ w |\psi_A|^4 + w_{AB} (2|\psi_A|^2 |\psi_B|^2 + |\psi_B|^4)
\right] \psi_A,
\label{eq:tdGL1}
\\
i\hbar\frac{\partial \psi_B}{\partial \tau} &=&
\left[ -\frac{\hbar^2}{2m} \nabla^2 - r({\bm x}) + u|\psi_B|^2 + u_{AB}|\psi_A|^2 
+ w |\psi_B|^4 + w_{AB} (2|\psi_A|^2 |\psi_B|^2 + |\psi_A|^4)
\right] \psi_B,
\label{eq:tdGL2}
\end{eqnarray}
which describe dynamics of the SF order-parameter fields.
From a mathematical perspective, they constitute a type of two-component NLSE with cubic and quintic nonlinearities. Thanks to the quintic terms, the SF states described by Eqs.~(\ref{eq:tdGL1}) and (\ref{eq:tdGL2}) have many peculiar properties that do not emerge in the GP equation~\cite{pitaevskii-03}.

Substituting $\psi_{\alpha}({\bm x},\tau) = \phi_{\alpha}({\bm x})$ into Eqs.~(\ref{eq:tdGL1}) and (\ref{eq:tdGL2}), the stationary part of the order parameter $\phi_{\alpha}({\bm x})$ obeys the time-independent GL equations,
\begin{eqnarray}
\left[ -\frac{\hbar^2}{2m} \nabla^2 - r({\bm x}) + u|\phi_A|^2 + u_{AB}|\phi_B|^2 
+ w |\phi_A|^4 + w_{AB} (2|\phi_A|^2 |\phi_B|^2 + |\phi_B|^4)
\right] \phi_A &=& 0,
\label{eq:tiGL1}
  \\
\left[ -\frac{\hbar^2}{2m} \nabla^2 - r({\bm x}) + u|\phi_B|^2 + u_{AB}|\phi_A|^2 
+ w |\phi_B|^4 + w_{AB} (2|\phi_A|^2 |\phi_B|^2 + |\phi_A|^4)
\right] \phi_B &=& 0.
\label{eq:tiGL2}
\end{eqnarray}

Next we consider small fluctuations from the stationary solution as 
\begin{eqnarray}
\psi_{\alpha}({\bm x},\tau) = 
\phi_{\alpha}({\bm x}) + \mathcal{U}_{\alpha}({\bm x})e^{-i\omega \tau} - \mathcal{V}_{\alpha}^{\ast}({\bm x})e^{i\omega^{\ast} \tau}, 
\end{eqnarray}
to obtain the Bogoliubov equations,
\begin{eqnarray}
\hat{M} {\bm U} = \hbar \omega {\bm U},
\label{eq:Bogo}
\end{eqnarray}
where
\begin{eqnarray}
{\bm U} &=& (\mathcal{U}_A, \mathcal{U}_B, \mathcal{V}_A, \mathcal{V}_B)^{\bf t},
\end{eqnarray}
and $\hat{M}$ is a $4\times4$ matrix whose elements are given by
\begin{eqnarray}
M_{11} &=& -M_{33} = 
-\frac{\hbar^2}{2m}\nabla^2 - r({\bm x}) + 2u |\phi_A|^2 + u_{AB} |\phi_B|^2 
+ 3w|\phi_A|^4 + w_{AB}(4|\phi_A|^2 |\phi_B|^2 + |\phi_B|^4),
\\
M_{22}&=& -M_{44} = 
-\frac{\hbar^2}{2m}\nabla^2 - r({\bm x}) + 2u |\phi_B|^2 + u_{AB} |\phi_A|^2 
+ 3w|\phi_B|^4 + w_{AB}(4|\phi_A|^2 |\phi_B|^2 + |\phi_A|^4),
\\
M_{12} &=& (M_{21})^{\ast} = u_{AB} \phi_A \phi_B^{\ast} 
+ 2w_{AB}\left(\phi_A^2 \phi_A^{\ast}\phi_B^{\ast} + \phi_A \phi_B (\phi_B^{\ast})^2\right),
\\
M_{13} &=& - (M_{31} )^{\ast} = 
- u \phi_A^2 - 2w \phi_A^3 \phi_A^{\ast} - 2 w_{AB} \phi_A^2 |\phi_B|^2,
\\
M_{14} &=& - (M_{41} )^{\ast} = -u_{AB}\phi_A \phi_B 
- 2 w_{AB} (\phi_A^2 \phi_A^{\ast} \phi_B +\phi_A \phi_B^2 \phi_B^{\ast} ),
\\
M_{23} &=& - (M_{32} )^{\ast} = -u_{AB}\phi_A \phi_B
- 2w_{AB}(\phi_A \phi_B^2 \phi_B^{\ast} +\phi_A^2 \phi_A^{\ast} \phi_B ), 
\\
M_{24} &=& - (M_{42} )^{\ast} =  - u \phi_B^2 - 2w \phi_B^3\phi_B^{\ast} 
- 2w_{AB} |\phi_A|^2\phi_B^2,
\\
M_{34} &=& (M_{43})^{\ast} = -u_{AB}\phi_A^{\ast}\phi_B 
- 2w_{AB} \left(\phi_A (\phi_A^{\ast})^2 \phi_B + \phi_A^{\ast} \phi_B^2 \phi_B^{\ast} \right).
\end{eqnarray}
Here, $\omega$ and ${\bm U}$ are the frequency and the amplitude of the normal mode of the SF order-parameter fields.
Stability of a stationary state is analyzed by solving the Bogoliubov equations.  
When normal modes with complex frequencies are present, the modes grow exponentially in time. This means that the state is dynamically unstable~\cite{pitaevskii-03}.

%%%%%%%%%%%%%%%%%
\section{Uniform solutions}\label{sec:phd}
%%%%%%%%%%%%%%%%%
In this section, we consider a uniform potential, $r({\bm x}) = r_0$, and a uniform solution,
%
%%
%\begin{eqnarray}
$\phi_A({\bm x})=\phi_B({\bm x})=\sqrt{n_0}$,
%\end{eqnarray}
%%
%
to analyze the SF-MI transition of the two-component BHM on the basis of the sixth-order GL theory. This theory is an established method for analyzing first-order transitions in general~\cite{binder-87}, and it has been applied to the same problem in previous works~\cite{yamamoto-13, kato-14}. We specifically aim to evaluate the state diagram of the SF along the axis of a dimensionless parameter, $u_{+}/(w_{+}n_0)$, which has not been addressed before. Here $u_{+} = u + u_{AB}$ and $w_{+} = w + 3w_{AB}$. We below assume that $w_{+} > 0$.

%%%%%%%%%%%%%%%%%%%%%%%%%%%%%%%%%%%
\begin{figure}[tb]
\includegraphics[scale=0.4]{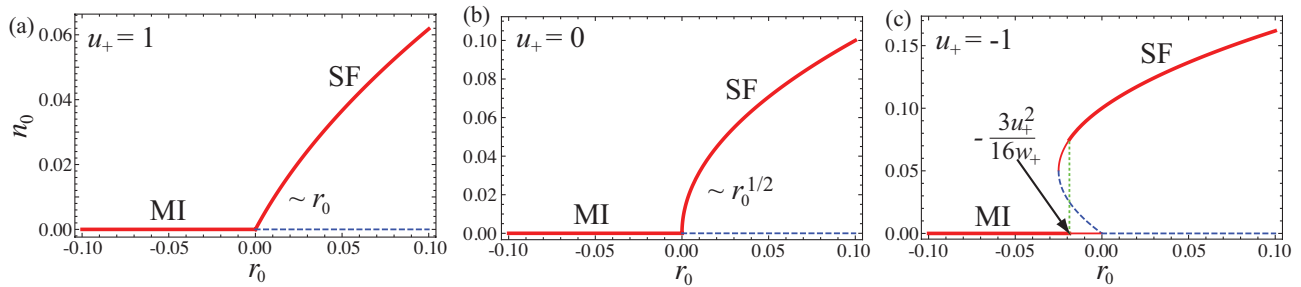}
\caption{\label{fig:dens}
(color online) Condensate density $n_0$ as a function of $r_0$, where we set $a=1$ and $w_{+}=10$. The thick-solid, thin-solid, and dashed lines represent the ground-state, metastable, and unstable solutions. The dotted line in (c) marks the first-order transition point.
}
\end{figure}
%%%%%%%%%%%%%%%%%%%%%%%%%%%%%%%%%%%

Substituting $\phi_A({\bm x})=\phi_B({\bm x})=\sqrt{n_0}$ into Eqs.~(\ref{eq:tiGL1}) and (\ref{eq:tiGL2}), one obtains
\begin{eqnarray}
\left(-r_0 + u_{+}n_0 + w_{+} n_0^2
\right) \sqrt{n_0} = 0.
\label{eq:homoGL}
\end{eqnarray}
There are two types of solution of Eq.~(\ref{eq:homoGL}). One is 
\begin{eqnarray}
n_0 = 0, 
\label{eq:miDens}
\end{eqnarray}
which corresponds to the MI state in the sense that the SF order parameters vanish.
The other type includes two SF states ($n_0>0$),
\begin{eqnarray}
n_0 &=& \frac{-u_{+} + \sqrt{u_{+}^2 + 4r_0 w_{+}}}{2w_{+}},
\label{eq:sfDens}
\\
n_0 &=& \frac{-u_{+} - \sqrt{u_{+}^2 + 4r_0 w_{+}}}{2w_{+}},
\label{eq:unDens}
\end{eqnarray}
which are the solutions of
\begin{eqnarray}
r_0=u_{+}n_0 + w_{+} n_0^2.
\label{eq:homoChem}
\end{eqnarray}

When $u_{+} \ge 0$, at which the transition is of second order, the two solutions of Eqs.~(\ref{eq:miDens}) and (\ref{eq:sfDens}) can satisfy the physical requirement $n_0\ge 0$.
When $r_0<0$, the MI state is the ground state and the SF state is forbidden by the condition $n_0\ge 0$. 
When $r_0>0$, the SF state is the ground state and the MI state is energetically unstable. 
As seen in Figs.~\ref{fig:dens}(a) and (b), when $r_0$ increases from the MI region, the second-order transition occurs at $r_0=0$, at which $n_0$ starts to grow from zero. When $u_{+} > 0$, the growth of $n_0$ behaves as $n_0 \sim r_0$ near the transition point~\cite{fisher-89}. This critical behavior can be captured even when the sixth-order terms are ignored. When $u_{+} = 0$, the transition is tricritical and $n_0 \sim r_0^{1/2}$~\cite{kato-14} as shown in Fig.~\ref{fig:dens}(b). 

%%%%%%%%%%%%%%%%%%%%%%%%%%%%%%%%%%%
\begin{figure}[tb]
\includegraphics[scale=0.5]{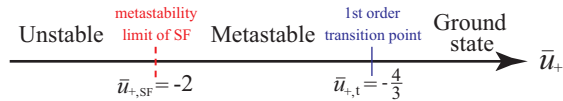}
\caption{\label{fig:PDu}
(color online) Mean-field state diagram for the SF described by the sixth-order GL action along the dimensionless parameter $\bar{u}_{+}\equiv u_{+}/(w_{+}n_0)$, where $w_{+}>0$.
}
\end{figure}
%%%%%%%%%%%%%%%%%%%%%%%%%%%%%%%%%%%

When $u_{+} < 0$, all the three solutions can satisfy $n_0\ge 0$. The solution of Eq.~(\ref{eq:unDens})
is present when $r_{\rm SF}\le r_0\le r_{\rm MI}$, and it corresponds to a dynamically unstable SF state, where $r_{\rm SF}= -u_+^2/(4w_+)$ and $r_{\rm MI}=0$ denote the metastability limit of the SF and MI states.
As illustrated in Fig.~\ref{fig:dens}(c), the MI state corresponding to the solution of Eq.~(\ref{eq:miDens}) is the ground state when $r_0< r_{\rm t}$, a metastable state when $r_{\rm t} < r_0 <r_{\rm MI}$, and an energetically unstable state when $r_0>r_{\rm MI}$. Here 
%
%%
%\begin{eqnarray}
$r_{\rm t} \equiv -\frac{3u_{+}^2}{16w_{+}}$
%\label{eq:rt}
%\end{eqnarray}
%%
%
denotes the first-order transition point that can be derived by means of Maxwell's construction. 
The SF state of Eq.~(\ref{eq:sfDens}) is the ground state when $r_{\rm t}<r_0$ and a metastable state when $r_{\rm SF}<r_0< r_{\rm t}$. The unstable SF state of Eq.~(\ref{eq:unDens}) connects the metastable SF with the metastable MI. Substituting $r_{0} = r_{\rm t}$ into Eq.~(\ref{eq:sfDens}), we evaluate the jump of $n_0$ at the transition point as $n_0 = -3u_{+}/(4w_{+})$. This means that when the dimensionless parameter $\bar{u}_{+} \equiv u_{+}/(w_{+}n_0)$ varies, the first-order transition occurs at $\bar{u}_{+}=-4/3\equiv \bar{u}_{+, {\rm t}}$. In a similar way, the metastability limit of the SF states is given by $\bar{u}_{+} = -2 \equiv \bar{u}_{+, {\rm SF}} $.
The state diagram of the SF along the axis of $\bar{u}_{+}$ is depicted in Fig.~\ref{fig:PDu}.

We next analyze normal modes of the uniform SF states, $\phi_A({\bm x})=\phi_B({\bm x})=\sqrt{n_0}$, in order to address dynamical stability. Substituting Eq.~(\ref{eq:homoChem}) into Eq.~(\ref{eq:Bogo}) and solving the eigenvalue problem, one obtains the dispersion relation,
\begin{eqnarray}
\hbar \omega_{\pm}({\bm p}) = 
\sqrt{\tilde{\epsilon}({\bm p})\left(\tilde{\epsilon}({\bm p}) + 2u_{\pm}n_0+4w_{\pm}n_0^2 \right)},
\label{eq:bogo-disp}
\end{eqnarray}
where ${\bm p}$ denotes the momentum of a normal mode, $\tilde{\epsilon}({\bm p}) \equiv \frac{p^2}{2m}$ the single-particle dispersion, $u_{-}=u-u_{AB}$, and $w_{-}=w-w_{AB}$. $\omega_{+}({\bm p})$ is the dispersion of in-phase modes while $\omega_{-}({\bm p})$ is that of out-of-phase modes. When the momentum is so small that $p \ll \sqrt{4m(u_{\pm}n_0 +2w_{\pm}n_0^2)}$, the normal modes take the form of phonon dispersion that is gapless and linear,
\begin{eqnarray}
\omega_{\pm}({\bm p}) \simeq c_{\pm}p,
\end{eqnarray}
where
\begin{eqnarray}
c_{\pm} \equiv \sqrt{\frac{u_{\pm}n_0 + 2w_{\pm}n_0^2}{m}}
\label{eq:ss}
\end{eqnarray}
is the sound speed.  Equation~(\ref{eq:ss}) tells us that when $u_{+} n_0 + 2w_{+}n_0^2<0$, or equivalently $\bar{u}_{+}<\bar{u}_{+,{\rm SF}}$, the in-phase modes at low momenta cause dynamical instability, leading to the collapse of the SF state. On the other hand, when $u_{-} n_0 + 2w_{-}n_0^2<0$, or equivalently $u_{-}/(w_{-}n_0) > -2$ for $w_{-}<0$, the out-of-phase modes at low momenta cause dynamical instability that results in the phase separation of the two components of the SF. Notice that $w_{-}<0$ is typically satisfied in the sixth-order GL action near the first-order SF-MI transition derived from the two-component BHM.

%%%%%%%%%%%%%%%
\section{Dark solitary waves}
\label{sec:soliton}
%%%%%%%%%%%%%%%
\subsection{One-dimensional treatment}
\begin{figure}[tb]
\includegraphics[scale=0.55]{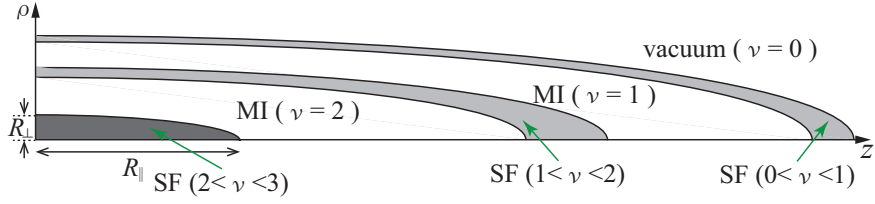}
\caption{\label{fig:shell}
(color online) Schematic picture of a spatial shell structure of the states in the two-component Bose-Hubbard system with a largely elongated trapping potential, where $\rho = \sqrt{x^2 + y^2}$.
}
\end{figure}

For a dark solitary wave to be long-lived, the motion of the SF states described by the sixth-order GL theory should be restricted only to one spatial direction, say, the $z$ direction. We explain below how one can prepare such a one-dimensional situation.
Let us consider the system of the two-component BHM of Eq.~(\ref{eq:hamiltonian}) at $d=3$ with a parabolic trapping potential,
\begin{eqnarray}
\epsilon_{\bm j} = \Omega_{\perp}(j_x^2 + j_y^2) + \Omega_{\parallel} j_z^2.
\end{eqnarray}
We assume that the potential is elongated towards the $z$ direction, i.e.,  $\Omega_{\parallel} \ll \Omega_{\perp}$, and that the filling factor at the trap center $\nu_{\rm ctr}$ satisfies $2<\nu_{\rm ctr}<3$. In such a setup, as indicated by the dark-shaded area in Fig.~\ref{fig:shell}, there is a region of the SF state with $2<\nu<3$ around the trap center, and it is surrounded by the shells of MI at $\nu=2$, SF at $1<\nu<2$, MI at $\nu=1$, SF at $0<\nu<1$, and vacuum ($\nu=0$). Such a shell structure of the phases has been observed in experiments with trapped single-component Bose gases in optical lattices~\cite{folling-06, campbell-06, gemelke-09}. We also assume the two conditions, namely $\nu_{\rm ctr}-2\ll 1$ and $R_{\perp}\gg a$, such that the SF state around the trap center can be well approximated by the sixth-order GL theory, where $R_{\perp}$ denotes the size of the SF region in the radial direction. The latter condition is necessary in order for the nature of the transition between MI at $\nu=2$ and SF at $2<\nu<3$ to be three dimensional. 

When the radial size $R_{\perp}$ is much smaller than the healing length $\xi$, the radial motion of the SF of $2<\nu<3$ can be regarded as frozen so that the order parameters is decomposed as~\cite{pitaevskii-03}
\begin{eqnarray}
\psi_{\alpha}({\bm x},\tau) = \psi_{\parallel,\alpha}(z, \tau) \phi_{\perp}(x, y),
\end{eqnarray}
where
\begin{eqnarray}
\phi_{\perp}(x,y) = \frac{1}{\sqrt{\pi}R_{\perp}}e^{-\frac{x^2+y^2}{2R_{\perp}^2}}.
\end{eqnarray}
Multiplying Eqs.~(\ref{eq:tdGL1}) and (\ref{eq:tdGL2}) by $\phi_{\perp}(x,y)$ and integrating them with respect to $x$ and $y$, one obtains the time-dependent GL equations in one dimension,
\begin{eqnarray}
i\hbar\frac{\partial \psi_{\parallel,A}}{\partial \tau}  &=& 
\left[ -\frac{\hbar^2}{2m} \frac{\partial^2}{\partial z^2} 
- r_{\parallel}(z) 
+ g|\psi_{\parallel,A}|^2 
+ g_{AB}|\psi_{\parallel,B}|^2 
+ f |\psi_{\parallel,A}|^4 
+ f_{AB} (2|\psi_{\parallel,A}|^2 |\psi_{\parallel,B}|^2 
+ |\psi_{\parallel,B}|^4)
\right] \psi_{\parallel,A},
\label{eq:1DtdGL1}
\\
i\hbar\frac{\partial \psi_{\parallel,B}}{\partial \tau} &=&
\left[ -\frac{\hbar^2}{2m} \frac{\partial^2}{\partial z^2} 
- r_{\parallel}(z) 
+ g|\psi_{\parallel,B}|^2 
+ g_{AB}|\psi_{\parallel,A}|^2 
+ f |\psi_{\parallel,B}|^4 
+ f_{AB} (2|\psi_{\parallel,A}|^2 |\psi_{\parallel,B}|^2 + |\psi_{\parallel,A}|^4)
\right] \psi_{\parallel,B},
\label{eq:1DtdGL2}
\end{eqnarray}
where $g=u/(2\pi R_{\perp}^2)$, 
$g_{AB}=u_{AB}/(2\pi R_{\perp}^2)$, 
$f=w/(3\pi^2 R_{\perp}^4)$, 
$f_{AB}=w_{AB}/(3\pi^2 R_{\perp}^4)$, 
$r_{\parallel}(z) = r_0 - \hbar\omega_{\perp} - \frac{1}{2}m\omega_{\parallel}^2 z^2$, 
and $\omega_{\parallel} = \sqrt{2C\Omega_{\parallel}/(ma^2)}$. 
Notice that the saddle-point approximation of the GL action, namely the mean-field equation of motion, is valid as long as the healing length is larger than the mean spacing of condensed particles in the SF state~\cite{pitaevskii-03}, i.e., $|\psi_{\parallel,\alpha}|^2\xi > 1$.
In the same way, the one-dimensional version of the time-independent GL equations (\ref{eq:tiGL1}) and (\ref{eq:tiGL2}), the Bogoliubov equations (\ref{eq:Bogo}), the state diagram of Fig.~\ref{fig:PDu}, and the dispersion relation of Eq.~(\ref{eq:bogo-disp}) are derived by the replacement of 
$\nabla^2 \rightarrow \frac{\partial^2}{\partial z^2}$,
$\phi_{\alpha}({\bm x})\rightarrow \phi_{\parallel,\alpha}(z)$,
${\bm U}({\bm x}) \rightarrow {\bm U}_{\parallel}(z)$,
$r({\bm x}) \rightarrow r_{\parallel}(z)$,
$u \rightarrow g$,
$u_{AB} \rightarrow g_{AB}$,
$w \rightarrow f$,
$w_{AB}\rightarrow f_{AB}$, and
$n_0 \rightarrow n_{\rm 1D}$.
Here we do not explicitly show those results to avoid redundancy.

%%%%%%%%%%%%%%%%%%%%%%%%%%%%%%%%%%%
%%%%%%%%%%%%%%%%%%%%%%%%%%%%%%%%%%%
%%%%%%%%%%%%%%%
\subsection{Analytical solutions}
%%%%%%%%%%%%%%%
We analytically solve the time-independent GL equations with a uniform potential to obtain dark solitary-wave solutions. For this purpose, assuming $\xi \ll R_{\parallel}$, we ignore the parabolic potential in the $z$ direction, i.e., we set $r_{\parallel}(z)=r_0-\hbar \omega_{\perp}\equiv r_{\rm 1D}$. We focus on a solution $\phi_{\parallel,A}(z) = \phi_{\parallel,B}(z) \equiv \phi(z)$, which is symmetric with respect to the exchange $A\leftrightarrow B$, and under this restriction the time-independent GP equations are simplified as
\begin{eqnarray}
\left( -\frac{\hbar^2}{2m} \frac{d^2}{dz^2} - r_{\rm 1D} + g_{+}|\phi|^2 + f_{+} |\phi|^4
\right) \phi = 0,
\label{eq:tidGL}
\end{eqnarray}
where $g_{+} = g+g_{AB}$ and $f_{+}=f+3f_{AB}$. Dark solitary-wave solutions of the single-component NLSE with cubic-quintic nonlinearity of Eq.~(\ref{eq:tidGL}) have been analyzed in previous studies~\cite{barashenkov-88, barashenkov-89, gagnon-89, kivshar-98, schurmann-00}. However, some specific properties of the dark solitary waves, including the size, the inertial mass, and the relation between the velocity and the phase jump, have not been explicitly described. Since the analytical solutions are necessary to obtain those properties, we start our calculations with a brief review of the derivation of the solutions. Notice that the solitary-wave solutions will be utilized also for investigating effects of a barrier potential in Sec.~\ref{sec:barrier}.

Let us find a solution with a single dark solitary wave under the boundary condition,
\begin{eqnarray}
\lim_{z\rightarrow \pm \infty}\phi(z) = 
\sqrt{n_{\rm 1D}}e^{i\left[q(z -z_{\rm s}) \pm \varphi/2\right]},
\label{eq:phiinf}
\end{eqnarray}
and
\begin{eqnarray}
|\phi(z)|<\sqrt{n_{\rm 1D}}, \,\,\, {\rm for} \,\,\, {\rm all} \,\,\, z,
\label{eq:phiall}
\end{eqnarray}
where $z_{\rm s}$ and $q$ denote the position of the soliton and the wave number of a supercurrent. $\varphi$ corresponds to the so-called phase jump of solitary-wave solutions. The boundary condition of Eq.~(\ref{eq:phiinf}) determines the relation between $r_{\rm 1D}$ and $n_{\rm 1D}$ as 
\begin{eqnarray}
r_{\rm 1D} = g_{+}n_{\rm 1D} + f_{+}n_{\rm 1D}^2 + \frac{\hbar^2 q^2}{2m},
\end{eqnarray}
which is an extension of Eq.~(\ref{eq:homoChem}) including the effect of the supercurrent.

We express the order parameter as $\phi(z)=\sqrt{n_{\rm 1D}}A(z)e^{iS(z)}$, where the amplitude $A(z)$ is real and positive, and the phase $S(z)$ is real. Substituting this into Eq.~(\ref{eq:tidGL}) leads to a set of equations as
\begin{eqnarray}
\left( -\frac{\hbar^2}{2m}\frac{d^2}{dz^2} + \frac{\hbar^2q^2}{2m}A^{-4}\right. \!\!\! &-& \!\!\!
\biggl. r_{\rm 1D} + g_{+}n_{\rm 1D} A^2 + f_{+}n_{\rm 1D}^2 A^4 \biggr) A
= 0, 
\label{eq:ample}
\\
%\end{eqnarray}
%\begin{eqnarray}
&&A^2 \frac{dS}{dz} = q. \label{eq:phase}
\end{eqnarray}
Equation~(\ref{eq:phase}) is the equation of continuity. Taking $f_{+}n_{\rm 1D}^2$ and $\xi_{f}\equiv \hbar/\sqrt{mf_{+}n_{\rm 1D}^2}$ as the units of the energy and length, Eqs.~(\ref{eq:ample}) and (\ref{eq:phase}) are rewritten in a dimensionless form,
\begin{eqnarray}
\left( -\frac{1}{2}\frac{d^2}{d\bar{z}^2} + \frac{\bar{q}^2}{2}A^{-4}\right. \!\!\! &-& \!\!\!
\biggl. \bar{r} + \bar{g}_{+} A^2 + A^4 \biggr) A
= 0, 
\label{eq:amplebar} 
\\
%\end{eqnarray}
%\begin{eqnarray}
&&A^2 \frac{dS}{d\bar{z}} = \bar{q}, 
\label{eq:phasebar}
\end{eqnarray}
where
\begin{eqnarray}
\bar{z} = \frac{z}{\xi_f}, \,\, 
\bar{r}=\frac{r_{\rm 1D}}{f_{+} n_{\rm 1D}^2}, \,\, 
\bar{g}_{+} = \frac{ g_{+} }{ f_{+} n_{\rm 1D}}, \,\, 
\bar{q}=q\xi_f.
%\,\, \bar{\tau}=\frac{\tau f_{+}n_{\rm 1D}^2}{\hbar}.
\end{eqnarray}

Multiplying Eq.~(\ref{eq:amplebar}) by $\frac{dA}{d\bar{z}}$ and integrating it with respect to $\bar{z}$, one obtains
\begin{eqnarray}
\left(\frac{dA}{d\bar{z}}\right)^2 = \frac{2}{3}A^{-2}(1-A^2)^2
\left(- \frac{3}{2}\bar{q}^2 + \gamma A^2 + A^4 \right),
\label{eq:middle}
\end{eqnarray}
where $\gamma \equiv 2 + 3\bar{g}_{+}/2$.
Integrating Eq.~(\ref{eq:middle}) again leads to an analytical expression of the amplitude,
\begin{eqnarray}
A(z) = \sqrt{\frac{\alpha_{+} + \alpha_{-}[\eta(z)]^2}{\beta_{+} - \beta_{-}[\eta(z)]^2}}
\label{eq:amplesol}
\end{eqnarray}
where
\begin{eqnarray}
\eta(z) &=& \tanh \left(\frac{z-z_{\rm s}}{\xi} \right),
\\
\alpha_{\pm} &=& \pm(-\gamma +3\bar{q}^2) + \sqrt{\gamma^2 + 6\bar{q}^2},
\\
\beta_{\pm} &=& 2 + \gamma \pm \sqrt{\gamma^2 + 6\bar{q}^2},
\end{eqnarray}
Notice that $\alpha_{\pm}\ge 0$ and $\beta_{\pm}\ge 0$ as long as $1+\gamma -\frac{3}{2}\bar{q}^2\ge 0$. 
From this solution, one immediately sees that the healing length is given by 
\begin{eqnarray}
\xi = \frac{\hbar}{\sqrt{m\Delta_0 - \hbar^2 q^2}},
\end{eqnarray}
where $\Delta_0 = g_{+}n_{\rm 1D} + 2f_{+}n_{\rm 1D}^2$.
Once the amplitude $A(z)$ is determined, one can calculate the phase $S(z)$ by converting Eq.~(\ref{eq:phase}) as
\begin{eqnarray}
S(z)-S(z_{\rm s}) = \int_{z_{\rm s}}^{z} dz \frac{q}{A^2}=
q(z-z_{\rm s})
+ {\rm sgn}(q)\arctan\left(\sqrt{\frac{\alpha_{-}}{\alpha_{+}}} \eta(z) \right),
\label{eq:phasesol}
\end{eqnarray}
where $S(z_{\rm s})$ is arbitrary due to the global $U(1)$ symmetry of the system, and we set $S(z_{\rm s}) = 0$.

From Eq.~(\ref{eq:amplesol}) and (\ref{eq:phasesol}), $\phi(z)$ is constructed as
\begin{eqnarray}
\frac{\phi(z)}{\sqrt{n_{\rm 1D}}} = A e^{iS} =
\frac{\sqrt{\alpha_{+}}+ i\,{\rm sgn}(q)
\sqrt{\alpha_{-}}\eta(z)}{\sqrt{\beta_{+}-\beta_{-}\left[\eta(z)\right]^2}}
e^{iq(z-z_{\rm s})}.
\label{eq:CWsol_tid}
\end{eqnarray}
This solution represents a dark solitary wave standing at $z=z_{\rm s}$ with a background supercurrent $v n_{\rm 1D}$ in the condensate, where $v\equiv \hbar q/m$ denotes the velocity of the supercurrent. Since we have assumed a uniform potential, the frame in which a solitary wave remains at rest can be transformed by the Galilean transformation into the one in which the solitary wave is moving at the velocity $-v$ in the static condensate,
\begin{eqnarray}
\frac{\psi'(z',\tau)}{\sqrt{n_{\rm 1D}}}=\frac{\phi(z)e^{-i q(z-z_{\rm s})}}{\sqrt{n_{\rm 1D}}}=
\frac{\sqrt{\alpha_{+}}+ i\,{\rm sgn}(q) 
\sqrt{\alpha_{-}}\eta(z'+v\tau)}
{\sqrt{\beta_{+}-\beta_{-}\left[\eta(z'+v\tau)\right]^2}},
\label{eq:CWsol}
\end{eqnarray}
where $z'\equiv z-v\tau$ and $\psi'(z',\tau)$ denote the position and the order parameter in the latter frame. The solution of Eq.~(\ref{eq:CWsol}) has been obtained in previous studies~\cite{barashenkov-88, barashenkov-89, gagnon-89, kivshar-98, schurmann-00}. 
When $\bar{g}_{+}\gg 1$, the solution of Eq.~(\ref{eq:CWsol}) is reduced to the celebrated dark-soliton solution of the GP equation~\cite{tsuzuki-71},
\begin{eqnarray}
\frac{\psi'(z',\tau)}{\sqrt{n_{\rm 1D}}}=\frac{v}{c_{g}}
+i\sqrt{1-\left(\frac{v}{c_{g}}\right)^2}
\tanh\left(\frac{z'+v\tau}{\xi_{g}}\right),
\end{eqnarray}
where $c_{g}\equiv \sqrt{g_{+}n_{\rm 1D}/m}$ and 
$\xi_{g}\equiv \hbar/\sqrt{mg_{+}n_{\rm 1D}-\hbar^2 q^2}$ denote the sound speed and the healing length at $f_{+}=0$.

%%%%%%%%%%%%%%%%%%%%%%%%%%%%%%%%%%%
\begin{figure}[tb]
\includegraphics[scale=0.55]{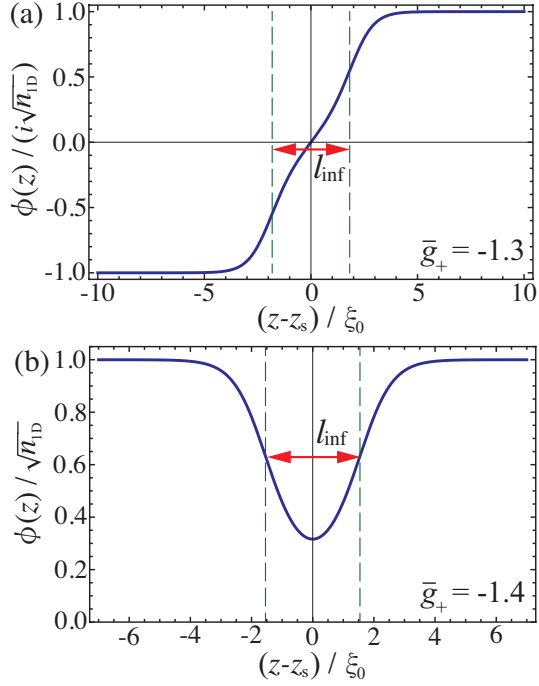}
\caption{\label{fig:solw}
(color online) Superfluid order parameter $\phi(z)$ at $\bar{g}_{+}=-1.3$ (a) and $\bar{g}_{+} = -1.4$ (b), where $q=0$ and $\xi_0 = \hbar/\sqrt{m\Delta_0}$. The dashed lines mark the inflection points.
}
\end{figure}
%%%%%%%%%%%%%%%%%%%%%%%%%%%%%%%%%%%
The structure of the solitary wave given by Eq.~(\ref{eq:CWsol}) substantially differs depending on whether the SF state is a ground state or a metastable state~\cite{barashenkov-88, barashenkov-89}. To see it clearly, we consider the case that $q=0$, in which the solitary wave is at rest. 
When $\gamma >0$, or equivalently $\bar{g}_+ > -4/3\equiv \bar{g}_{+,{\rm t}}$, the SF state is a ground state and the solitary-wave solution of Eq.~(\ref{eq:CWsol}) is simplified as
\begin{eqnarray}
\frac{\phi(z)}{\sqrt{n_{\rm 1D}}} = \frac{i\sqrt{\gamma} 
\eta(z)}{\sqrt{1+\gamma - [\eta(z)]^2}}.
\label{eq:phi_dark}
\end{eqnarray}
In Fig.~\ref{fig:solw}(a), we plot $\phi(z)/(i\sqrt{n_{\rm 1D}})$ at $q=0$ and $\bar{g}_{+}=-1.3$. This solitary wave has a $\pi$ phase jump at $z=z_{\rm s}$, where the amplitude vanishes, and in this sense it is a topological excitation. The topological property is the same as that for the black soliton of the GP equation~\cite{pitaevskii-03}. 

When $-1 <\gamma <0$, or equivalently $\bar{g}_{+,{\rm SF}}<\bar{g}_{+}<\bar{g}_{+,{\rm t}}$, the SF state is metastable and Eq.~(\ref{eq:CWsol}) is reduced to
\begin{eqnarray}
\frac{\phi(z)}{\sqrt{n_{\rm 1D}}} = \sqrt{ \frac{ -\gamma}{1 - (1+\gamma) [\eta(z)]^2} },
\label{eq:phi_bld}
\end{eqnarray}
where $\bar{g}_{+,{\rm SF}}= -2$.
In Fig.~\ref{fig:solw}(b), we plot $\phi(z)/\sqrt{n_{\rm 1D}}$ at $q=0$ and $\bar{g}_{+}=-1.4$. It is worth noting that there is no phase jump in this solitary wave in contrast to the standard solitary wave with a $\pi$ phase jump. Thus, the dark solitary wave of a metastable SF is a non-topological excitation and is often called a bubble~\cite{barashenkov-88,barashenkov-89}.

When $\gamma<-1$, or equivalently $\bar{g}_{+}<\bar{g}_{+,{\rm SF}}$, the integration of Eq.~(\ref{eq:middle}) to Eq.~(\ref{eq:amplesol}) fails and the solution of Eq.~(\ref{eq:CWsol}) is irrelevant. This is associated with the fact that in this region of $\gamma$ the SF state is dynamically unstable towards collapse.

%%%%%%%%%%%%%%%%%%%%%%%%%%%%%%%%%%%
\begin{figure}[tb]
\includegraphics[scale=0.55]{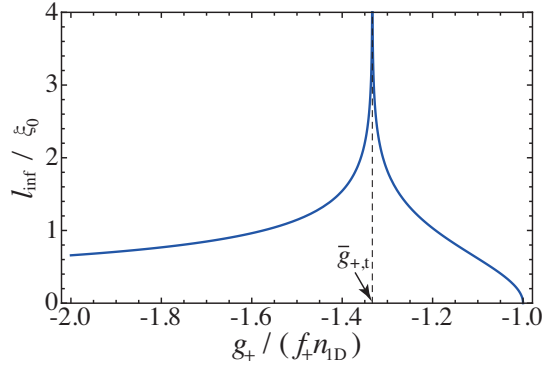}
\caption{\label{fig:lsol}
(color online) $l_{\rm inf}$ versus $\bar{g}_{+}$ at $q=0$. The dashed line marks the first-order transition point.
}
\end{figure}
%%%%%%%%%%%%%%%%%%%%%%%%%%%%%%%%%%%
In order to explain how the solitary wave transforms from the $\pi$-jumped shape to the bubble, we focus on a certain length scale $l_{\rm inf}$ that quantifies the size of a solitary wave near the transition point. When $\gamma < 1/2$, or equivalently $\bar{g}_{+}<-1$, there emerges two additional inflection points in $\phi(z)$, which are marked as the cross points with the green dashed lines in Fig.~\ref{fig:solw}. From the condition, $\frac{d^2\phi}{dz^2}=0$, the inflection points are determined, and we define $l_{\rm inf}$ as the distance between the two, which is given at $q=0$ by
\begin{eqnarray}
\frac{l_{\rm inf}}{\xi_0} = \left\{ \begin{array}{cc}
2\,{\rm arctanh} \left( 1-2 \gamma \right)^{\frac{1}{2}}
& {\rm if} \,\,\, \bar{g}_{+,{\rm t}}<\bar{g}_{+}<-1
\\
2\,{\rm arctanh} \left( 1-2\gamma \right)^{-\frac{1}{2}}
& {\rm if} \,\,\, \bar{g}_{+,{\rm SF}}<\bar{g}_{+}<\bar{g}_{+,{\rm t}}
\end{array}\right.
,
\end{eqnarray}
where $\xi_0 \equiv \hbar/\sqrt{m\Delta_0}$ is the healing length at $q=0$.
In Fig.~\ref{fig:lsol}, we plot $l_{\rm inf}$ as a function of $\bar{g}_{+}$. When the transition point $\bar{g}_{+} = \bar{g}_{+,{\rm t}}$ is approached, $l_{\rm inf}$ diverges as $l_{\rm inf}/\xi_0 \simeq -\ln|\bar{g}_{+}-\bar{g}_{+,{\rm t}}|$. This means that there is no solution satisfying the boundary condition of Eq.~(\ref{eq:phiinf}) at the transition point and that this singularity separates the two different types of solitary wave. 

%%%%%%%%%%%%%%%%%%%%%%%%%%%%%%%%%%%
\begin{figure}[tb]
\includegraphics[scale=0.55]{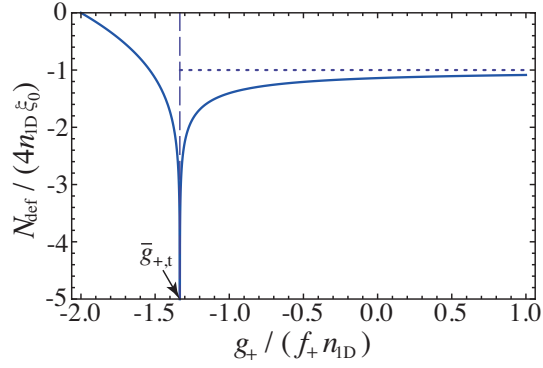}
\caption{\label{fig:deficit}
(color online) Deficit of condensed particles $N_{\rm def}$ at $q=0$ as a function of $\bar{g}_{+}$. The dashed line marks the first-order transition point. The dotted line represents the asymptote of $N_{\rm def}$ at $\bar{g}_{+}\rightarrow \infty$.
}
\end{figure}
%%%%%%%%%%%%%%%%%%%%%%%%%%%%%%%%%%%
The same divergent behavior is seen also in the number~\cite{barashenkov-88},
\begin{eqnarray}
N_{\rm def}=\sum_{\alpha}\int_{-\infty}^{\infty}dz\left[ |\psi_{\parallel,\alpha}(z,\tau)|^2 - n_{\rm 1D} \right],
\label{eq:nsol_def}
\end{eqnarray}
which represents the deficit of condensed particles due to the density reduction around $z=z_{\rm s}$. Substituting Eqs.~(\ref{eq:phi_dark}) and (\ref{eq:phi_bld}) into Eq.~(\ref{eq:nsol_def}), one obtains the deficit number at $q=0$,
\begin{eqnarray}
\frac{N_{\rm def}}{n_{\rm 1D}\xi_0} = \left\{ \begin{array}{cc}
-4\sqrt{1+\gamma} \,{\rm arctanh} \left(1+\gamma \right)^{-\frac{1}{2}},
& {\rm if} \,\,\, \bar{g}_{+}>\bar{g}_{+,{\rm t}}
\\
-4\sqrt{1+\gamma}\,{\rm arctanh} \left( 1 + \gamma \right)^{\frac{1}{2}},
& {\rm if} \,\,\, \bar{g}_{+,{\rm SF}}<\bar{g}_{+}<\bar{g}_{+,{\rm t}}
\end{array}\right.
.
\end{eqnarray}
In Fig.~\ref{fig:deficit}, we plot $N_{\rm def}$ as a function of $\bar{g}_{+}$. Near the transition point, since the deficit corresponds to a hole with the density $-2n_{\rm 1D}$ and the size $l_{\rm inf}$, the deficit number diverges as $N_{\rm def} \simeq -2n_{\rm 1D}l_{\rm inf}\simeq 2n_{\rm 1D}\xi_0 \ln|\bar{g}_{+}-\bar{g}_{+,{\rm t}}|$.

%%%%%%%%%%%%%%%%%%%%%%%%%%%%%%%%%%%
\begin{figure}[tb]
\includegraphics[scale=0.55]{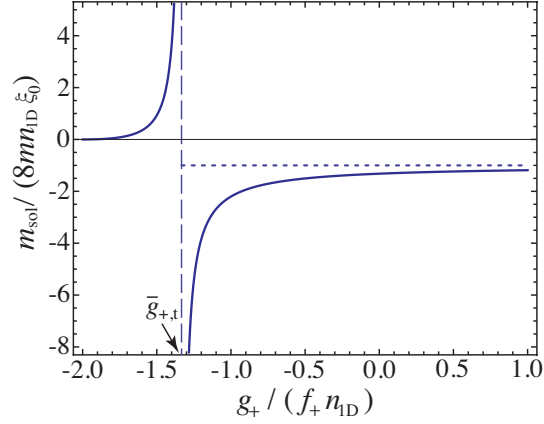}
\caption{\label{fig:mass}
(color online) Inertial mass of the dark solitary wave $m_{\rm sol}$ at $q=0$ as a function of $\bar{g}_{+}$. The dashed line marks the first-order transition point. The dotted line represents the asymptote of $m_{\rm sol}$ at $\bar{g}_{+}\rightarrow \infty$.
}
\end{figure}
%%%%%%%%%%%%%%%%%%%%%%%%%%%%%%%%%%%
%%%%%%%%%%%%%%%%%%%%%%%%%%%%%%%%%%%
The divergence of $l_{\rm inf}$ is interesting in the sense that it indicates that some properties of nonlinear excitations exhibit criticality associated with the first-order transition, which does not appear either in any linear excitations or in any thermodynamic quantities of the uniform SF state. However, since the divergence is logarithmic, it seems rather hard to observe in experiments. We therefore show that the inertial mass of the solitary wave $m_{\rm sol}$ exhibits a stronger divergence near the first-order transition. The inertial mass is defined as
\begin{eqnarray}
m_{\rm sol} = 2\frac{\partial}{\partial (v^2)}\Delta E,
\label{eq:mass_def}
\end{eqnarray}
where
\begin{eqnarray}
\Delta E \equiv E_{\rm sol} - E_{\rm 0}
\label{eq:Esurplus}
\end{eqnarray}
represents the surplus of the energy of the SF state due to the presence of the dark solitary wave~\cite{scott-11, pitaevskii-14}. The energy of the SF state for a given $\psi_{\parallel,\alpha}(z,\tau)$ is
\begin{eqnarray}
E=\int_{-\infty}^{\infty}\!\!\!&dz&\!\!\!
\left[
\sum_{\alpha}\left(
\left|\frac{\partial \psi_{\parallel,\alpha}}{\partial z}\right|^2
-r_{\parallel}\left|\psi_{\parallel,\alpha}\right|^2 
+ \frac{g}{2}\left|\psi_{\parallel,\alpha}\right|^4
+ \frac{f}{3}\left|\psi_{\parallel,\alpha}\right|^6
\right)\right. \nonumber \\
&&
\left.
+ g_{AB} \left|\psi_{\parallel,A}\right|^2 \left|\psi_{\parallel,B}\right|^2
+ f_{AB} \left(\left|\psi_{\parallel,A}\right|^4 \left|\psi_{\parallel,B}\right|^2
+ \left|\psi_{\parallel,A}\right|^2 \left|\psi_{\parallel,B}\right|^4 \right)
\right].
\label{eq:energy}
\end{eqnarray}
$E_{\rm sol}$ and $E_{0}$ in Eq.~(\ref{eq:Esurplus}) mean the energies for the solitary-wave solution of Eq.~(\ref{eq:CWsol}) and the uniform solution $\psi_{\parallel,\alpha} = \sqrt{n_{\rm 1D}}$, respectively. The analytical expression of $\Delta E$ is given by
\begin{eqnarray}
\Delta E = \frac{f_{+}n_{\rm 1D}^3\xi_f}{\sqrt{6}}
\left[
(4\gamma + \gamma^2 - 6\bar{q}^2){\rm arctanh} 
\left(
\sqrt{\frac{\beta_{-}}{\beta_{+}}}
\right)
+\sqrt{1+\gamma-\frac{3}{2}\bar{q}^2}(2-\gamma)
\right].
\label{eq:Eresult}
\end{eqnarray}
Substituting Eq.~(\ref{eq:Eresult}) into Eq.~(\ref{eq:mass_def}), one obtains the inertial mass, and its analytical expression for $v\ll c_{+}$ is written as
\begin{eqnarray}
\frac{m_{\rm sol}}{m n_{\rm 1D}\xi_0} = \left\{ \begin{array}{cc}
-4\sqrt{1+\gamma}\,{\rm arctanh} \left(1+\gamma \right)^{-\frac{1}{2}} - 4- \frac{4}{\gamma},
& {\rm if} \,\,\, \bar{g}_{+} > \bar{g}_{+,{\rm t}}
\\
-4\sqrt{1+\gamma}\,{\rm arctanh} \left(1+\gamma \right)^{\frac{1}{2}} - 4 - \frac{4}{\gamma},
& {\rm if} \,\,\, \bar{g}_{+,{\rm SF}}<\bar{g}_{+}<\bar{g}_{+,{\rm t}}
\end{array}\right.
.
\end{eqnarray}
In Fig.~\ref{fig:mass}, we plot $m_{\rm sol}$ at $v\rightarrow 0$ as a function of $\bar{g}_{+}$. When the transition point is approached, the inertial mass diverges as 
\begin{eqnarray}
\frac{m_{\rm sol}}{mn_{\rm 1D}\xi_0}\simeq -\frac{8}{3(\bar{g}_{+}-\bar{g}_{+,{\rm t}})}.
\end{eqnarray}
Since the divergence of this form is stronger than the logarithmic one, the inertial mass is advantageous over the other quantities $l_{\rm inf}$ and $N_{\rm def}$ for experimental observation of the divergent behavior.

%%%%%%%%%%%%%%%%%%%%%%%%%%%%%%%%%%%
\begin{figure}[tb]
\includegraphics[scale=0.55]{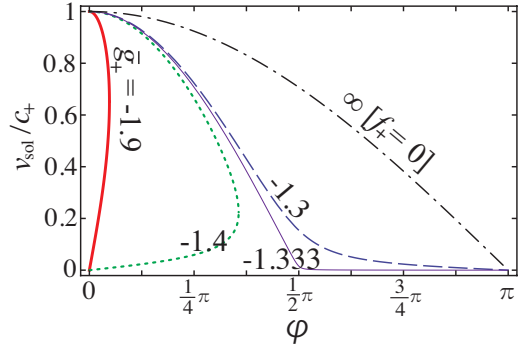}
\caption{\label{fig:vph}
(color online) Soliton velocity $v$ as a function of the phase jump $\varphi$.
}
\end{figure}
%%%%%%%%%%%%%%%%%%%%%%%%%%%%%%%%%%%
A possible way for observing the divergence of the inertial mass is to measure the relation between the velocity $v$ and the phase jump $\varphi$ of the solitary wave, because the inertial mass is directly connected with $\frac{d\varphi}{dv}$ through the equation~\cite{scott-11, pitaevskii-14},
\begin{eqnarray}
m_{\rm sol} = mN_{\rm def} + 2\hbar n_{\rm 1D} \frac{d\varphi}{dv}.
\end{eqnarray}
To derive the relation between $v$ and $\varphi$, let us express the phase $S(z)$ of Eq.~(\ref{eq:phasesol}) at a distance far from $z=z_{\rm s}$,
\begin{eqnarray}
\lim_{z \rightarrow \pm \infty}S(z) = q(z-z_{\rm s}) 
\pm {\rm sgn}(q) \arctan
\left( \sqrt{\frac{\alpha_{-}}{\alpha_{+}}} \right).
\label{eq:phaseinf}
\end{eqnarray}
Comparing Eq.~(\ref{eq:phaseinf}) to the boundary condition of Eq.~(\ref{eq:phiinf}), one immediately obtains
\begin{eqnarray}
\varphi = 
2\,{\rm sgn}(q)\arctan
\left( \sqrt{\frac{\alpha_{-}}{\alpha_{+}}} \right).
\label{eq:phiofv}
\end{eqnarray}
Since in cold-atom experiments the velocity $v$ is measured after the phase jump $\varphi$ is created with the use of phase-imprinting techniques~\cite{becker-08, yefsah-13}, it is convenient to express $v$ as a function of $\varphi$ by solving Eq.~(\ref{eq:phiofv}) with respect to $v=\hbar q/m$.
When the SF state is a ground state ($\bar{g}_{+}>\bar{g}_{+,{\rm t}}$), one obtains
\begin{eqnarray}
\frac{v}{c_{+}} = \left\{ \begin{array}{cc}
\sqrt{\frac{\cos^2\varphi + \gamma  
+ \cos \varphi \sqrt{\cos^2\varphi +2\gamma +\gamma^2 } }
{2(1 + \gamma)}}, 
& {\rm if} \,\,\, 0<\varphi<\pi \\
-\sqrt{\frac{ \cos^2\varphi  + \gamma
+ \cos \varphi \sqrt{ \cos^2\varphi +2\gamma + \gamma^2} }
{2(1 + \gamma)}}, 
& {\rm if} \,\,\, -\pi<\varphi<0
\end{array}\right.
.
\label{eq:vofphi_1}
\end{eqnarray}
In the limit of $\bar{g}_{+} \rightarrow \infty$, Eq.~(\ref{eq:vofphi_1}) reduces to the velocity-phase relation of the dark soliton of the GP equation~\cite{pitaevskii-03},
\begin{eqnarray}
\frac{v}{c_{g}} = \left\{ \begin{array}{cc}
\cos\frac{\varphi}{2}, 
& {\rm if} \,\,\, 0<\varphi<\pi \\
-\cos\frac{\varphi}{2}, 
& {\rm if} \,\,\, -\pi<\varphi<0
\end{array}\right.
.
\label{eq:vofphi_2}
\end{eqnarray}
When the SF state is metastable ($\bar{g}_{+,{\rm SF}}<\bar{g}_{+}<\bar{g}_{+,{\rm t}}$),
\begin{eqnarray}
\frac{v}{c_{+}} = \left\{ \begin{array}{cc}
\sqrt{\frac{ \cos^2\varphi  + \gamma
\pm \cos \varphi \sqrt{ \cos^2\varphi +2\gamma + \gamma^2} }
{2(1+\gamma)}}, 
& {\rm if} \,\,\, 0<\varphi<\varphi_{\rm max} \\
-\sqrt{\frac{\cos^2\varphi + \gamma
\pm \cos \varphi \sqrt{\cos^2\varphi +2\gamma  + \gamma^2} }
{2(1 + \gamma)}}, 
& {\rm if} \,\,\, -\varphi_{\rm max}<\varphi<0
\end{array}\right.
,
\end{eqnarray}
where $\varphi_{\rm max} = \arcsin(1+\gamma)$.
In Fig.~\ref{fig:vph}, we plot $v$ as a function of $\varphi$ for several values of $\bar{g}_{+}$. When $\bar{g}_{+}>\bar{g}_{+,{\rm t}}$, $v$ is a one-valued function of $\varphi$. As $\bar{g}_{+}$ decreases towards the transition point, the value of $v$ in the region of $\pi/2 < |\varphi| < \pi$ is significantly suppressed, which is indicated by the dashed and thin-solid lines in Fig.~\ref{fig:vph}. When the transition point is crossed, $v$ abruptly changes to a two-valued function of $\varphi$ and the domain of $\varphi$ is shrunk to $0<|\varphi|<\varphi_{\rm max}$ (see the dotted and thick-solid line in Fig.~\ref{fig:vph}).

Differentiating Eq.~(\ref{eq:phiofv}) with respect to $v$ leads to 
\begin{eqnarray}
\frac{d\varphi}{dv} = - 
\frac{-2(\gamma+\gamma^2+3\bar{q}^2)\sqrt{1+\gamma}}
{c_{+}(\gamma^2+6\bar{q}^2)
\sqrt{1+\gamma-\frac{3}{2}q^2}},
\end{eqnarray}
and taking the limit of $v\rightarrow 0$, one obtains
\begin{eqnarray}
\frac{d\varphi}{dv} = - \frac{2}{c_{+}}\left(1+\frac{1}{\gamma}\right).
\end{eqnarray}
Thus, the quantity $\frac{d\varphi}{dv}$ exhibits the divergence of the form that 
$\frac{d\varphi}{dv} \sim (\bar{g}_{+,{\rm t}}-\bar{g}_{+})^{-1}$, reflecting the divergence of the inertial mass. Hence, the divergence of the inertial mass may be observed through the measurement of the relation between $v$ and $\varphi$.

%%%%%%%%%%%%%%%%%%%%%%%%%%%%%%%%%%%
\begin{figure}[tb]
\includegraphics[scale=0.55]{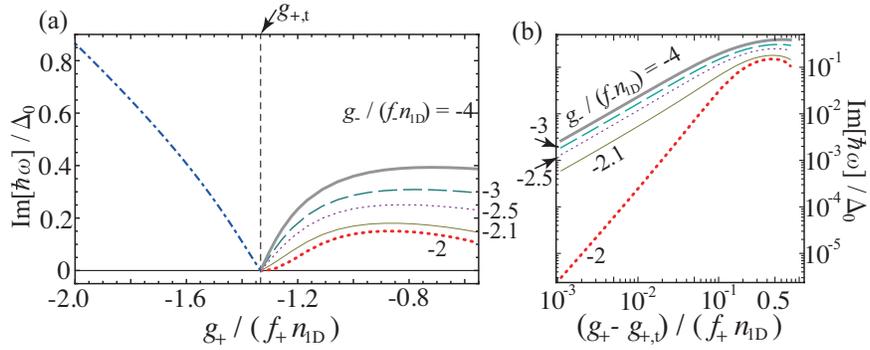}
\caption{\label{fig:imome}
(color online) Imaginary part of the frequency of the most dominant unstable mode in the dark solitary-wave solutions as a function of $\bar{g}_{+}$. In (b), data are shown in a log-log scale.
}
\end{figure}
%%%%%%%%%%%%%%%%%%%%%%%%%%%%%%%%%%%

\subsection{Stability analysis}
For experimental observation of the dark solitary waves, it is important to be aware of whether or not the solitary waves are dynamically stable, and if not, whether or not its lifetime is long enough for measurement of the solitary waves. To answer these questions, we perform a linear stability analysis by numerically solving the Bogoliubov equations (\ref{eq:Bogo}) with the solitary wave solutions of Eqs.~(\ref{eq:phi_dark}) and (\ref{eq:phi_bld}) at $q=0$. Unfortunately, we find that the solitary waves are dynamically unstable at any values of $\bar{g}_{+}$ and $g_{-}/(f_{-}n_{\rm 1D})$, where $g_{-} = g - g_{AB}$ and $f_{-} = f - f_{AB}$. In Fig.~\ref{fig:imome}, we plot the imaginary part of the frequency ${\rm Im}[\omega]$ of the most dominant unstable mode, whose inverse corresponds to the lifetime of the solitary wave. In the bubble-like solitary wave of a metastable SF state ($\bar{g}_{+,{\rm SF}}<\bar{g}_{+}<\bar{g}_{+,{\rm t}}$), the unstable mode is an in-phase mode, and the lifetime does not depend on $g_{-}/(f_{-}n_{\rm 1D})$. Notice that the dynamical instability of a bubble-like solitary wave has been pointed out in previous studies~\cite{barashenkov-88,barashenkov-89}. On the other hand, in the standard dark-soliton of a ground-state SF ($\bar{g}_{+}>\bar{g}_{+,{\rm t}}$), the unstable mode is an out-of-phase mode, which is specific to the two-component system, and the lifetime is longer for smaller $g_{-}/(f_{-}n_{\rm 1D})$. In both cases, when the first-order transition point ($\bar{g}_{+}=\bar{g}_{+,{\rm t}}$) is approached, ${\rm Im}[\omega]$ decreases towards zero, meaning that the lifetime becomes infinitely long. Thus, the lifetime of the solitary wave can be sufficiently long for experimental realization near the transition point.

%%%%%%%%%%%%%
\section{Barrier potential}
\label{sec:barrier}
In Sec.~\ref{sec:soliton}, we have seen that some properties of dark solitary waves exhibit critical behaviors in the vicinity of the first-order transition point. However, since the solitary waves are dynamically unstable excited states, it is better if there is a certain system that exhibits the same criticality in a stable equilibrium state.
In this section, we investigate effects of a barrier potential on the SF state described by the sixth-order GL theory to predict that the SF state with a barrier potential exemplifies such a system. Specifically, we consider a barrier potential of a $\delta$-functional form located at $z=0$,
\begin{eqnarray}
r_{\parallel}(z) = r_{\rm 1D} - V\delta(z), 
\end{eqnarray}
where $V>0$ is a barrier strength. Notice that while we continue to adopt the one-dimensional situation for consistency with Sec.~\ref{sec:soliton}, discussions in this section are valid in 3D systems with a barrier potential in the axial direction and a uniform potential in the radial direction, namely
\begin{eqnarray}
r({\bm x}) = r_0 - V\delta(z).
\end{eqnarray}

\begin{figure}[tb]
\includegraphics[scale=0.55]{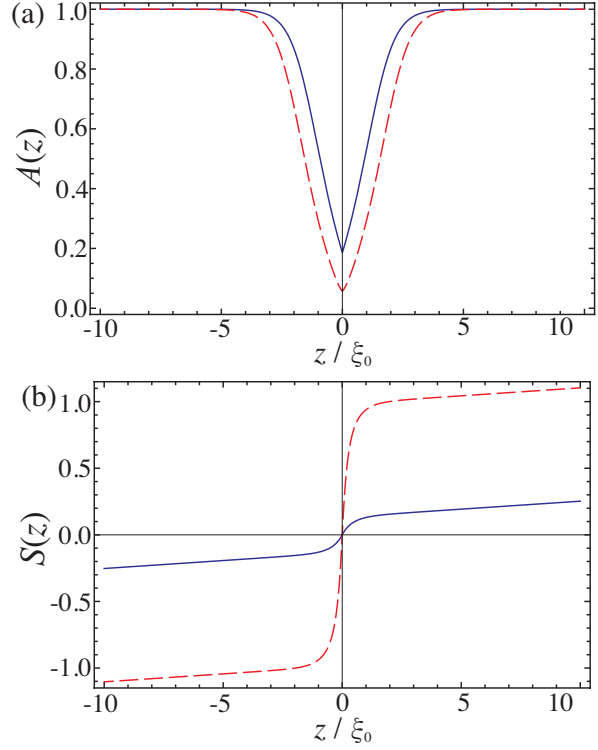}
\caption{\label{fig:AandS}
(color online) Amplitude $A(z)$ and phase $S(z)$ of the order parameter field $\phi(z)$ in the presence of a barrier potential. We set $\bar{g}_{+} = -1.3$, $\bar{q}=0.01$, and $\bar{V}=1.2$. The solid and dashed lines represent the stable and unstable solutions.
}
\end{figure}

We assume that the condensate possesses a supercurrent $Q$ that flows through the barrier potential. In such a situation, the order parameters satisfy the boundary conditions of Eqs.~(\ref{eq:phiinf}) and (\ref{eq:phiall}). Since there is no external potential except at $z=0$, one can obtain the solution of the time-independent GL equations by using almost the same procedure described in the previous section,
\begin{eqnarray}
A(z) = \sqrt{\frac{\alpha_{+} + \alpha_{-}[\tilde{\eta}(z)]^2}{\beta_{+} - \beta_{-}[\tilde{\eta}(z)]^2}},
\label{eq:amplesol_imp}
\end{eqnarray}
\begin{eqnarray}
S(z) = 
qz
+ {\rm sgn}(qz)\left[
\arctan\left(\sqrt{\frac{\alpha_{-}}{\alpha_{+}}} \tilde{\eta}(z) \right)
-
\theta_0
\right],
\label{eq:phasesol_imp}
\end{eqnarray}
and 
\begin{eqnarray}
\frac{\phi(z)}{\sqrt{n_{\rm 1D}}} = 
e^{i\left(qz - {\rm sgn}(qz)\theta_0\right)}
\frac{
\sqrt{\alpha_{+}} + i \, {\rm sgn} (qz) \sqrt{\alpha_{-}}\tilde{\eta}(z)
}
{\sqrt{\beta_{+} - \beta_{-} \left[ \tilde{\eta}(z) \right]^2}},
\label{eq:phisol_imp}
\end{eqnarray}
where
\begin{eqnarray}
\tilde{\eta}(z) = \tanh\left(\frac{|z|+z_0}{\xi}\right),
\end{eqnarray}
and
\begin{eqnarray}
\theta_0 = \arctan \left( \sqrt{\frac{\alpha_{-}}{\alpha_{+}}} \tilde{\eta}(0) \right).
\end{eqnarray}
In Eq.~(\ref{eq:phasesol_imp}), we have set $S(0) = 0$. The only one major difference from the previous case is that the constant $z_0$ has to be determined by the boundary condition at $z=0$,
\begin{eqnarray}
\left.\frac{dA}{dz}\right|_{z=+0} &=& 
\left.\frac{dA}{dz}\right|_{z=-0}+\frac{2mV}{\hbar^2} A(z=0).
\label{eq:bc_imp}
\end{eqnarray}
Notice that the solution of Eqs.~(\ref{eq:amplesol_imp}) and (\ref{eq:phasesol_imp}) obviously satisfies the other boundary conditions at $z=0$,
\begin{eqnarray}
A(z=+0) &=& A(z=-0),
\label{eq:bc1}
\\
S(z=+0) &=& S(z=-0), 
\label{eq:bc2}
\\
\left.\frac{dS}{dz}\right|_{z=+0} &=& \left.\frac{dS}{dz}\right|_{z=-0},
\label{eq:bc3}
\end{eqnarray}
Substituting Eq.~(\ref{eq:amplesol_imp}) into Eq.~(\ref{eq:bc_imp}) leads to
\begin{eqnarray}
4\sqrt{\frac{2}{3}\left(1+\gamma -\frac{3}{2}\bar{q}^2  \right)^3(\gamma^2 +6\bar{q}^2)}
\left(\tilde{\eta}(0)-\left[\tilde{\eta}(0)\right]^3\right)
=\bar{V}
\left(\alpha_{+} + \alpha_{-}\left[ \tilde{\eta}(0) \right]^2\right)
\left(\beta_{+} - \beta_{-}\left[ \tilde{\eta}(0) \right]^2\right).
\label{eq:z0det}
\end{eqnarray}
where $\bar{V} = V/(f_{+}n_{\rm 1D}^2\xi_{f})$.
Solving Eq.~(\ref{eq:z0det}) for given values of $\bar{g}_{+}$, $\bar{q}$, and $\bar{V}$, one determines $z_0$ (or equivalently $\tilde{\eta}(0)$), which fixes the solution of Eq.~(\ref{eq:phisol_imp}). In Figs.~\ref{fig:AandS}(a) and (b), we plot $A(z)$ and $S(z)$ for $\bar{g}_{+} = -1.3$, $\bar{q}=0.01$, and $\bar{V}=1.2$ as an example. Typically, there are two solutions. The one with a smaller phase jump has a smaller energy and is stable. In the limit of $V\rightarrow 0$, it becomes a solution with a flat density, i.e., $\phi(z) = \sqrt{n_{\rm 1D}}e^{iqz}$. The other with a larger phase jump is dynamically unstable and becomes the dark solitary-wave solution of Eq.~(\ref{eq:CWsol_tid}) at $V\rightarrow 0$.

%%%%%%%%%%%%%%%%%%%%%%%%%%%%%%%%%%%
%%%%%%%%%%%%%%%%%%%%%%%%%%%%%%%%%%%%

\subsection{Critical barrier strength}
%%%%%%%%%%%%%%%%%%%%%%%%%%%%%%%%%%%
\begin{figure}[tb]
\includegraphics[scale=0.55]{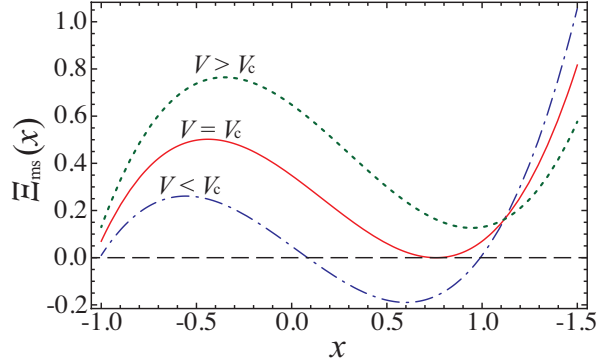}
\caption{\label{fig:funcofx}
(color online) Function $\Xi_{\rm ms}(x)$ of Eq.~(\ref{eq:xims}) at $\bar{g}_{+} = -1.467$. The dotted, solid, and dash-dotted lines correspond to $\bar{V}=\bar{V}_{\rm c}+0.3$, $\bar{V}_{\rm c}$, and $\bar{V}_{\rm c}-0.3$, where $\bar{V}_{\rm c}=0.3487$.
}
\end{figure}
\begin{figure}[tb]
\includegraphics[scale=0.55]{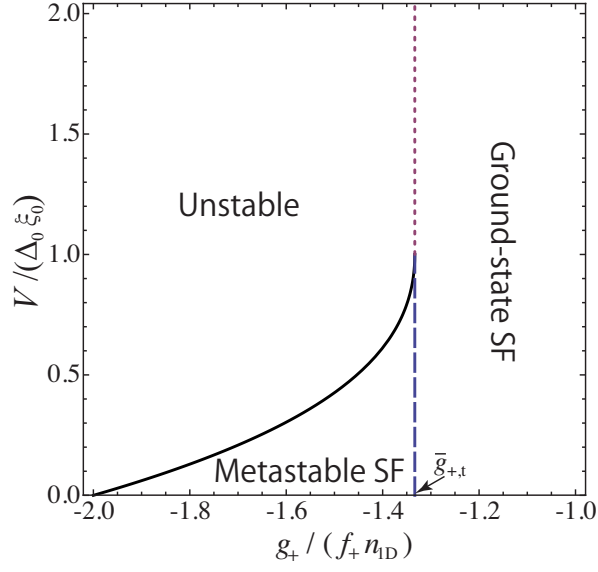}
\caption{\label{fig:vc}
(color online) State diagram in the $(\bar{g}_{+}, \tilde{V})$ plane at $q=0$. The solid line represents the critical barrier strength of Eq.~(\ref{eq:vc}).
}
\end{figure}

%%%%%%%%%%%%%%%%%%%%%%%%%%%%%%%%%%%%
We consider the case that the condensate is at rest in order to show that a barrier potential destroys a metastable SF state when $V$ exceeds a certain critical value. When the SF state is metastable 
($\bar{g}_{+,{\rm SF}}<\bar{g}_{+}<\bar{g}_{+,{\rm t}}$) and $q=0$, Eq.~(\ref{eq:z0det}) is simplified as
\begin{eqnarray}
\Xi_{\rm ms}\left(\tilde{\eta}(0)\right) = 0,
\label{eq:xizero}
\end{eqnarray}
where
\begin{eqnarray}
\Xi_{\rm ms}\left(x\right) = \bar{V} - \sqrt{\frac{2}{3}(1+\gamma)^{3}} x - \bar{V}(1+\gamma) x^2 + \sqrt{\frac{2}{3}(1+\gamma)^{3}}x^3.
\label{eq:xims}
\end{eqnarray}
In Fig.~\ref{fig:funcofx}, we plot the function $\Xi(x)$ for different three values of $\bar{V}$. As indicated by the dash-dotted line in Fig.~\ref{fig:funcofx}, when $V$ is smaller than a certain threshold value, say $V_{\rm c}$, Eq.~(\ref{eq:xizero}) has two solutions in the range of $0\le \tilde{\eta}(0) \le 1$. In contrast, when $V>V_{\rm c}$, there is no solution of Eq.~(\ref{eq:xizero}) at $0\le \tilde{\eta}(0) \le 1$ (see the dotted line in Fig.~\ref{fig:funcofx}). This happens because of the bubble-like structure of the dark solitary wave in the metastable SF state. Since the density minimum of the bubble-like solution of Eq.~(\ref{eq:phi_bld}) is always finite, i.e., $A(z_{\rm s})>0$, the metastable SF state can not hold a sufficiently deep density-dip that is favored for a strong barrier potential.  From Eq.~(\ref{eq:xizero}), one can derive an analytical expression of the critical barrier strength,
\begin{eqnarray}
\tilde{V}_{\rm c} &=& 
\sqrt{\frac{8-20\gamma-\gamma^2 - \sqrt{-\gamma(8-\gamma)^3}}{8(1+\gamma)}}
\label{eq:vc}
\\
&\simeq&
\left\{ \begin{array}{cc}
1 - \sqrt{3\left(\bar{g}_{+,{\rm t}} - \bar{g}_{+}\right)}, & 
{\rm if} \,\,\, \bar{g}_{+} \simeq \bar{g}_{+,{\rm t}} 
\\
\frac{1}{\sqrt{3}}\left(\bar{g}_{+}-\bar{g}_{+,{\rm SF}}\right), & {\rm if} \,\,\, \bar{g}_{+} \simeq \bar{g}_{+,{\rm SF}} 
\end{array}
\right.
,
\end{eqnarray}
where $\tilde{V}_{\rm c}=V_{\rm c}/(\Delta_0 \xi_0)$. In Fig.~\ref{fig:vc}, we plot the critical barrier strength of Eq.~(\ref{eq:vc}) by the solid line.
Notice that when the $\delta$-function potential is attractive, i.e., $V<0$, there is not such a threshold value of the potential strength for the presence of a stable SF solution.

%%%%%%%%%%%%%%%%%%%%%%%%%%%%%%%%%%%
%%%%%%%%%%%%%%%%%%%%%%%%%%%%%%%%%%%%
 
\begin{figure}[tb]
\includegraphics[scale=0.55]{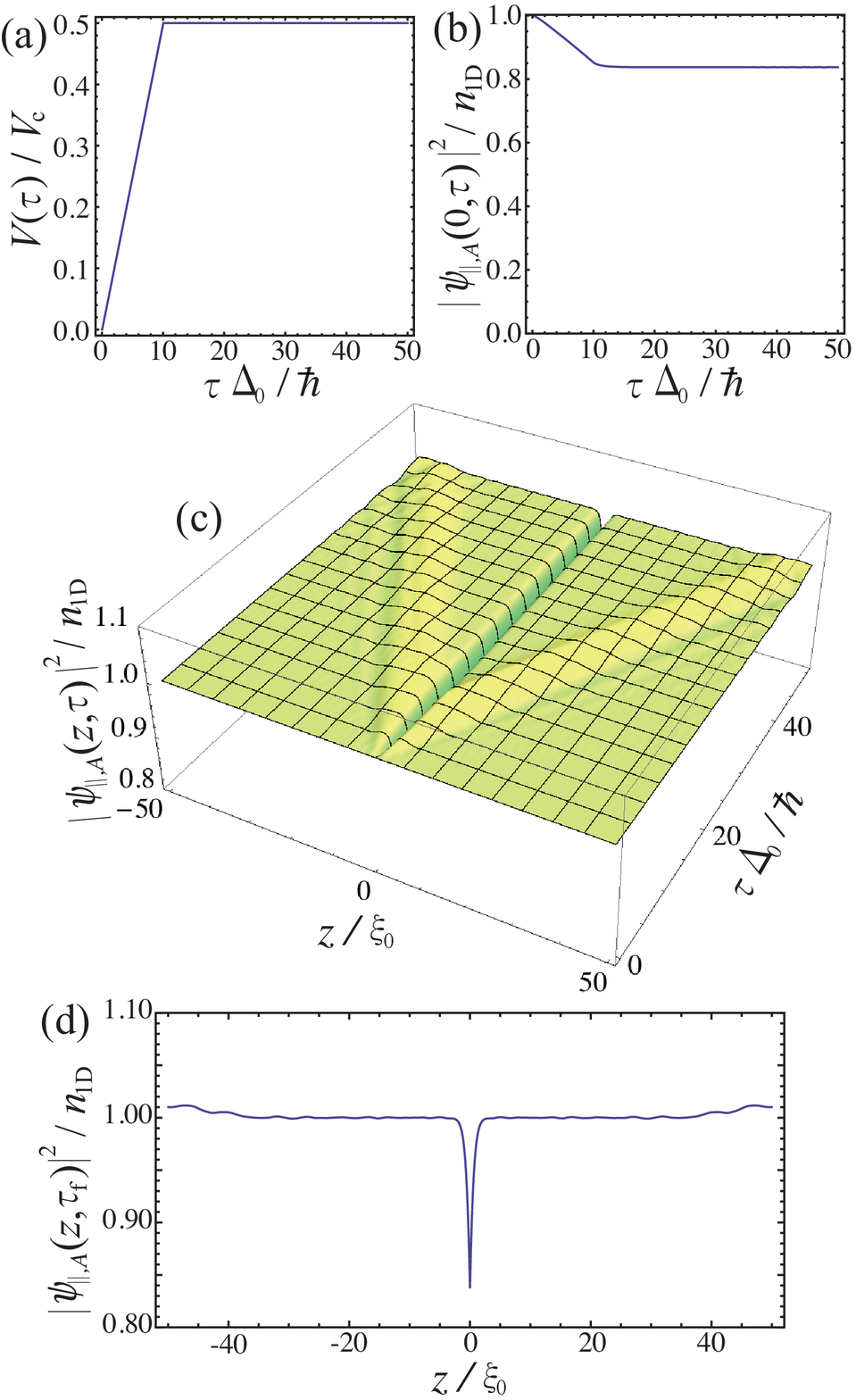}
\caption{\label{fig:dynamLow}
(color online) Dynamics of the SF state in response to the linear ramp of the barrier potential. We set $q=0$, $\bar{g}_{+}=-1.6$, $\bar{g}_{-} = 6$, $V_{\rm max} = 0.5 V_{\rm c}$, and $\tau_{\rm rp}=10\hbar / \Delta_0$. In (a), (b), and (c), the time evolutions of the barrier potential, the density at $z=0$, and the density distribution are depicted. In (d), the density distribution at $\tau=\tau_{\rm f}= 50\hbar / \Delta_{0}$ is plotted.
}
\end{figure}

\begin{figure}[tb]
\includegraphics[scale=0.55]{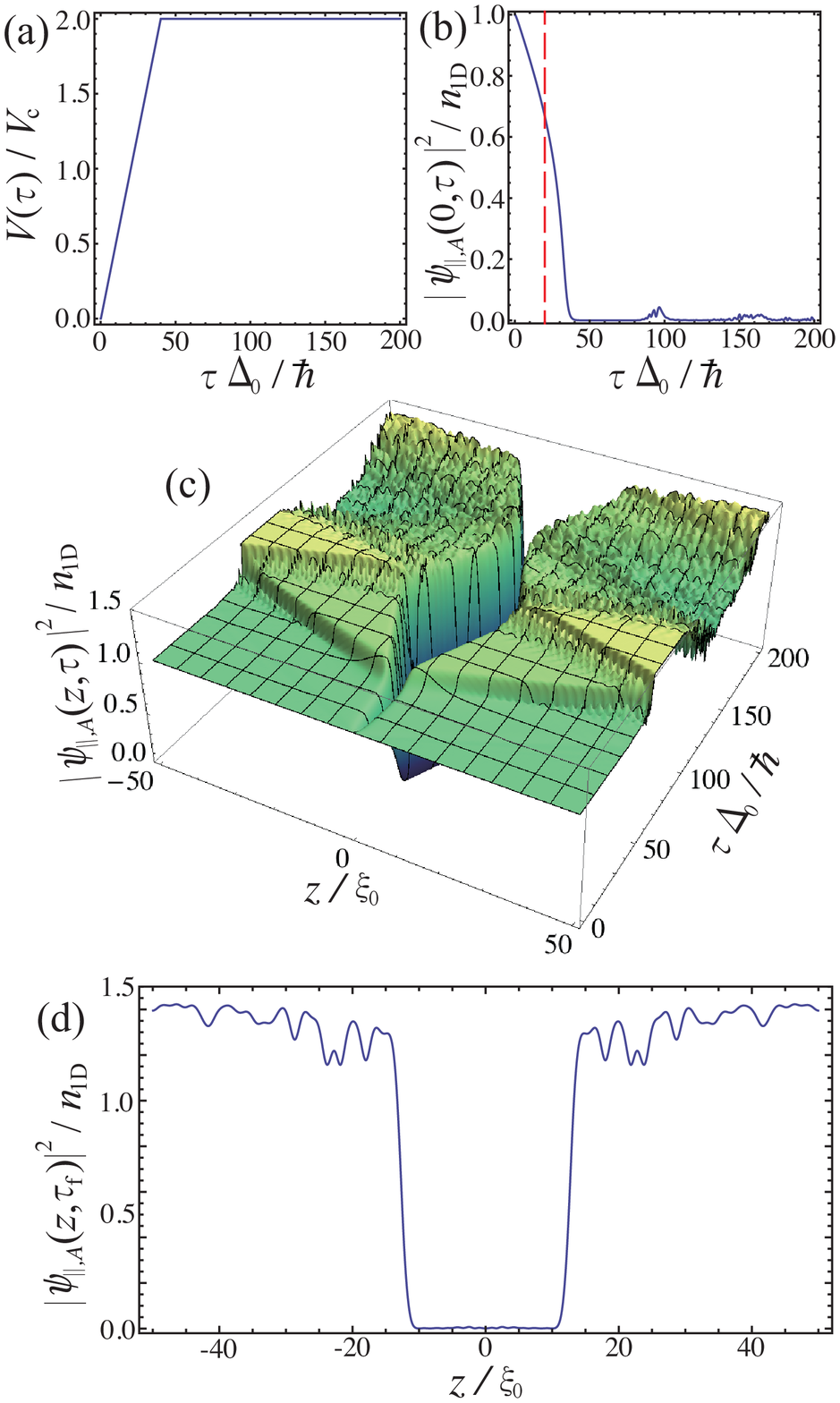}
\caption{\label{fig:dynamHigh}
(color online) Dynamics of the SF state in response to the linear ramp of the barrier potential. We set $q=0$, $\bar{g}_{+}=-1.6$, $\bar{g}_{-} = 6$, $V_{\rm max} = 2 V_{\rm c}$, and $\tau_{\rm rp}=40\hbar / \Delta_{0}$. In (a), (b), and (c), the time evolutions of the barrier potential, the density at $z=0$, and the density distribution are depicted. In (d), the density distribution at $\tau=\tau_{\rm f}= 200\hbar / \Delta_{0}$ is plotted.
}
\end{figure}

The absence of solutions at $V>V_{\rm c}$ indicates that the metastable SF state may be destroyed by the strong barrier. To corroborate this, we compute real-time dynamics of the metastable SF state subject to a linear ramp of the barrier potential,
\begin{eqnarray}
V(\tau) =
\left\{
\begin{array}{cc}
V_{\rm max} \frac{\tau}{\tau_{\rm rp}}, & \,\, 0\le \tau < \tau_{\rm rp}
\\
V_{\rm max}, & \tau > \tau_{\rm rp}
\end{array}
\right.
,
\end{eqnarray}
by solving the time-dependent GL equations of Eqs.~(\ref{eq:tdGL1}) and (\ref{eq:tdGL2}). Here $V_{\rm max}$ and $\tau_{\rm rp}$ denote the maximum value of $V(\tau)$ and the ramp time of the barrier. Specific forms of $V(\tau)$ used in our calculations are illustrated in Figs.~\ref{fig:dynamLow}(a) and \ref{fig:dynamHigh}(a). In the calculations below, we take the periodic boundary condition, $\bar{g}_{+}=-1.6$, $g_{-}/(w_{-}n_{\rm 1D}) = -6$, and $L=100\xi_{0}$, where $L$ denotes the system size in the $z$ direction. At this value of $\bar{g}_{+}$, the critical barrier strength is $\tilde{V}_{\rm c} = 0.3036$.

Let us first analyze the case that $V_{\rm max} < V_{\rm c}$. In Fig.~\ref{fig:dynamLow}, we show the results for $V_{\rm max} = 0.5 V_{\rm c}$ and $\tau_{\rm rp} = 10 \hbar/\Delta_0$. As seen in Figs.~\ref{fig:dynamLow}(b), (c), and (d), the density around the barrier is suppressed in response to the repulsion by the barrier, and the density distribution remains almost steady after $\tau=\tau_{\rm rp}$. This exemplifies the fact that the metastable SF state is compatible with a barrier potential as long as $V<V_{\rm c}$.

We next consider the case that $V_{\rm max} > V_{\rm c}$. In Fig.~\ref{fig:dynamHigh}, we show the results for $V_{\rm max} = 2 V_{\rm c}$ and $\tau_{\rm rp} = 40 \hbar/\Delta_0$, where $\tau_{\rm rp}$ is taken such that the ramp rate is the same as the previous case. In Fig.~\ref{fig:dynamHigh}(b) and (c), before $V(\tau)$ reaches $V_{\rm c}$, the density around the barrier exhibits a nearly linear suppression as in the previous case. However, after $V_{\rm c}$ is exceeded, the density suppression is dramatically accelerated until the density around the barrier vanishes. The vanishing density means that the barrier potential of $V>V_{\rm c}$ completely disrupts the SF as illustrated in Fig.~\ref{fig:dynamHigh}(d).

%%%%%%%%%%%%%%%%%%%%%%%%%%%%%%%%%%%
%%%%%%%%%%%%%%%%%%%%%%%%%%%%%%%%%%%%
%%%%%%%%%%%%%%%%%%%%%%%%%%%%%%%%%%%
%%%%%%%%%%%%%%%%%%%%%%%%%%%%%%%%%%%%

When the SF state is a ground state ($\bar{g}_{+}>\bar{g}_{+,{\rm t}}$), the SF state is stable for any strength of the repulsive barrier and there is no critical barrier strength. In this case, Eq.~(\ref{eq:z0det}) at $q=0$ is reduced to 
\begin{eqnarray}
\Xi_{\rm gs}\left(\tilde{\eta}(0)\right) = 0,
\label{eq:xidash}
\end{eqnarray}
where
\begin{eqnarray}
\Xi_{\rm gs}\left(x\right) = \sqrt{\frac{2}{3}(1+\gamma)^{3}}x - \bar{V} (1+\gamma) x^2 - \sqrt{\frac{2}{3}(1+\gamma)^{3}} x^3 
+ \bar{V}x^4.
\end{eqnarray}
Equation~(\ref{eq:xidash}) has a trivial solution $\tilde{\eta}(0) = 0$, which implies $z_{0} = 0$. Such a solution is equivalent to the dark solitary-wave solution of Eq.~(\ref{eq:phi_dark}) standing at $z=0$. Moreover, there is always another solution that satisfies $0\le \tilde{\eta}(0) \le 1$. For instance, when $\bar{V}\gg 1$, the latter solution is $\tilde{\eta}(0) \simeq 1/\tilde{V}$, where $\tilde{V} = V/(\Delta_0 \xi_0)$, and it obviously exists even up to the limit of $V\rightarrow \infty$.

%%%%%%%%%%%%%%%%%%%%%%%%%
\subsection{Barrier-induced criticality}
%%%%%%%%%%%%%%%%%%%%%%%%%
The fact that the critical barrier strength terminates at a finite value ($\tilde{V}_{\rm c}= 1$) in the limit of $\bar{g}_{+}\nearrow \bar{g}_{+,{\rm t}}$ means that the line of the metastability limit represented by the solid line in Fig.~\ref{fig:vc} switches to the line of $\bar{g}_{+} = \bar{g}_{+,{\rm t}}$ above $\tilde{V}=1$ as indicated by the dotted line in Fig.~\ref{fig:vc}. At the metastability limit resulting from the barrier, the size $l_{\rm inf}$ of the density dip around the barrier diverges logarithmically in the same way as the divergence of $l_{\rm inf}$ of the dark solitary waves seen in Sec.~\ref{sec:soliton}.
We recall that $l_{\rm inf}$ is defined as the distance between the two inflection points of $A(z)$, and the one for $\bar{g}_{+} > \bar{g}_{+,{\rm t}}$ in the presence of the repulsive barrier is given by
\begin{eqnarray}
\frac{l_{\rm inf}}{\xi_{0}} = 2{\rm arctanh}(1-2\gamma)^{\frac{1}{2}} -\frac{z_0}{\xi_0}.
\label{eq:linf_v}
\end{eqnarray}
To gain simple analytical insights, let us consider the region of $\tilde{V} \gg 1$, where $z_0 / \xi_0 \simeq 1/\tilde{V} \ll 1$ and the second term in the right-hand side of Eq.~(\ref{eq:linf_v}) can be ignored. In this case, it is easy to see that when $r_{\rm 1D}$ is varied from the ground-state SF side to the transition point $r_{\rm 1D, t}$, the density-dip size diverges logarithmically as $l_{\rm inf}\simeq - \xi_0 \ln \left( \frac{r_{\rm 1D}}{r_{\rm 1D, t}} - 1 \right)$. This logarithmic divergence survives even out of the region of $\tilde{V} \gg 1$ as long as $\tilde{V} \ge 1$.

At a glance it seems that the first-order SF-MI transition shifts to a second-order one associated with the switch of the metastability limit line. However, this is only true in the strict limit of $L\rightarrow \infty$, and the transition is weak first order at a large but finite system~\cite{lipowsky-84, sornette-85, lipowsky-87}. To explain this, we evaluate the energy of the SF state in the presence of a barrier potential and compare it with the energy of the MI state that is zero within the mean-field approximation. Substituting Eq.~(\ref{eq:phisol_imp}) into Eq.~(\ref{eq:energy}) at $\bar{g}_{+} > \bar{g}_{+,{\rm t}}$, $q=0$, and $\tilde{V}\gg 1$, one obtains
\begin{eqnarray}
E \simeq f_{+}n_{\rm 1D}^3L\left[
-\frac{2}{3}\gamma 
+ \frac{\xi_{f}}{\sqrt{6}L}\left(
(4\gamma + \gamma^2){\rm arctanh}\left(1+\gamma\right)^{-\frac{1}{2}}
+ \sqrt{1+\gamma}(2-\gamma)
\right)
\right].
\end{eqnarray}
The condition $E=0$ leads to the first-order transition point in the presence of the barrier potential,
\begin{eqnarray}
\bar{g}_{+,\ast} \simeq \bar{g}_{+,{\rm t}} + \frac{2\xi_0}{3L}.
\label{eq:shift_gt}
\end{eqnarray}
Thus, the shift of the transition point, $\bar{g}_{+,\ast}-\bar{g}_{+,{\rm t}}$, is on the order of $\xi_0/L$, which vanishes in the strictly thermodynamic limit. 
From Eq.~(\ref{eq:shift_gt}), the size of the density dip at the shifted transition point $\bar{g}_{+} = \bar{g}_{+,\ast}$ is given by 
$l_{\rm inf} \simeq \xi_0 \ln \frac{L}{\xi_0}$.
This means that the first-order transition precedes the divergence of $l_{\rm inf}$, and the critical regime near the metastability limit can not be reached if one looks only at ground states. However, since the SF state is stable anyway up to the metastability limit, one can approach the critical region by starting from a ground-state SF and changing the parameter $\bar{g}_{+}$ towards the metastability limit.  

The divergence of the density-dip size leads to the emergence of criticality in thermodynamics quantities, such as the averaged density $n_{\rm ave}= N_{\rm tot}/L$ and the compressibility $\kappa=\frac{1}{n_{\rm ave}}\frac{\partial n_{\rm ave}}{\partial \mu}\propto \frac{\partial n_{\rm ave}}{\partial r_{\rm 1D}}$.  Here the total number of condensed particles $N_{\rm tot}$ is given by
\begin{eqnarray}
N_{\rm tot} = \sum_{\alpha}\int dz |\psi_{\parallel,\alpha}(z,\tau)|^2.  
\label{eq:ntot_def}
\end{eqnarray}
Substituting Eq.~(\ref{eq:phisol_imp}) at $\gamma>0$ and $q=0$ into Eq.~(\ref{eq:ntot_def}), one obtains
\begin{eqnarray}
n_{\rm ave} &=& 2n_{\rm 1D} \left[1 -  
2\frac{\xi_0}{L}\sqrt{1+\gamma}\left( {\rm arctanh}\left(\frac{1}{\sqrt{1+\gamma}}\right) 
- {\rm arctanh}\left(\frac{\tilde{\eta}(0)}{\sqrt{1+\gamma}}\right) \right) 
\right],
\\
&\simeq& 2n_{\rm 1D} \left[ 1 + \frac{\xi_0}{L}\ln\left(\frac{ r_{\rm 1D} } { r_{\rm 1D,t}}-1\right) \right].
\label{eq:nave_lim}
\end{eqnarray}
The compressibility exhibits a stronger signature of the criticality as $\kappa \propto \frac{\xi_0}{L} \left( \frac{r_{\rm 1D}}{r_{\rm 1D, t}}-1 \right)^{-1}$. However, the divergent behavior of $\kappa$ is prominent only at a tiny region, 
$\frac{r_{\rm 1D}}{r_{\rm 1D, t}}-1 \lesssim \xi_0/L$, and such precise control of the parameter is unrealistic in experiments.
As we will see below, a critical behavior of the critical current emerges in a much wider range of the parameter.

%%%%%%%%%%%%%%%%%%%%%%%%%%%%%%%%%%%
\subsection{Current-phase characteristics}
%%%%%%%%%%%%%%%%%%%%%%%%%%%%%%%%%%%
\begin{figure}[tb]
\includegraphics[scale=0.55]{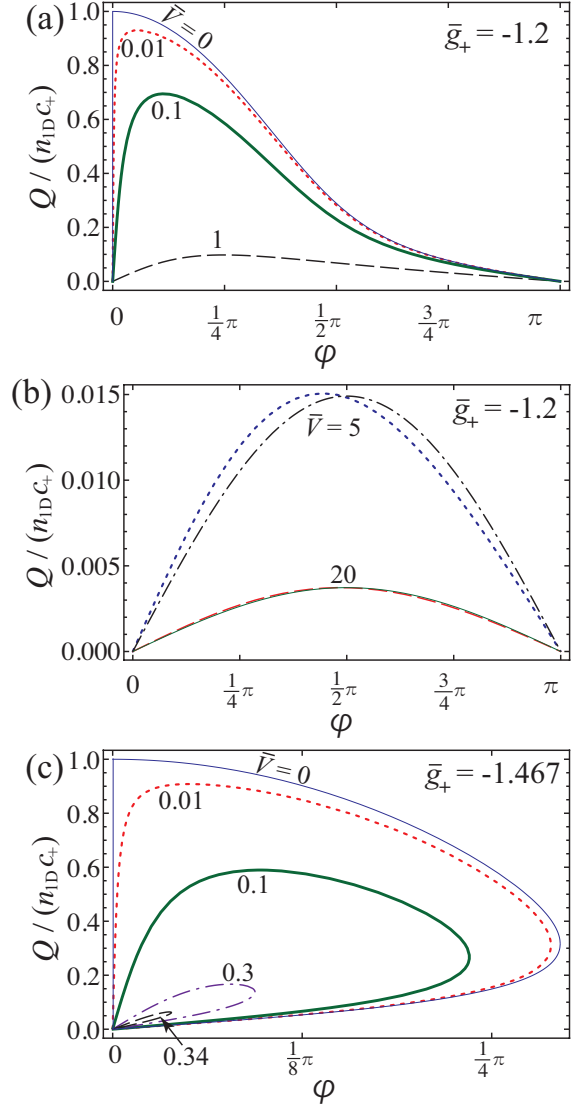}
\caption{\label{fig:qphi}
(color online) Current $Q$ as a function of the phase jump $\varphi$ at $\bar{g}_{+}=-1.2$ (a) and (b), and $\bar{g}_{+}=-1.467$ (c). The dash-dotted ($\bar{V}=5$) and thin-solid ($\bar{V}=20$) lines represent the Josephson relation of Eq.~(\ref{eq:joseph}).
}
\end{figure}
%%%%%%%%%%%%%%%%%%%%%%%%%%%%%%%%%%%%
Applying the boundary condition of Eq.~(\ref{eq:phiinf}) to the solution of Eq.~(\ref{eq:phisol_imp}), one obtains the relation between the phase jump $\varphi$ and the current $Q=\hbar q/m$,
\begin{eqnarray}
\varphi = 2 \, {\rm sgn}(q) \left[
\arctan\left(\sqrt{\frac{\alpha_{-}}{\alpha_{+}}}  \right)
-
\theta_0
%\arctan\left(\sqrt{\frac{\alpha_{-}}{\alpha_{+}}} \tilde{\eta}(0) \right)
\right].
\end{eqnarray}
In Figs.~\ref{fig:qphi}(a) and (b), we depict the current-phase relation at $\bar{g}_{+} = -1.2>\bar{g}_{+,{\rm t}}$.
%, where the SF state at $q=\varphi = 0$ is a ground state. 
When $\bar{V}$ increases, it asymptotically approaches the Josephson relation~\cite{josephson-62},
\begin{eqnarray}
\frac{Q}{n_{\rm 1D} c_{+}} \equiv \tilde{Q} \simeq
\frac{\gamma}{2(1+\gamma)\tilde{V}}\sin \varphi,
\label{eq:joseph}
\end{eqnarray}
which is indicated by the dash-dotted ($\bar{V} = 5$) and thin-solid ($\bar{V}=20$) lines in Fig.~\ref{fig:qphi}(b). In the region of $\bar{g}_{+,{\rm SF}}<\bar{g}_{+}<\bar{g}_{+,{\rm t}}$, 
%where the SF state at $q=\varphi = 0$ is metastable, 
topology of the current-phase relation is different. As seen in Fig.~\ref{fig:qphi}(c), $Q$ is a two-valued function of $\varphi$, which forms a loop structure. This peculiar behavior stems from the fact that the dark solitary wave at $q=0$ is non-topological. When $\bar{V}$ increases, the loop shrinks, and it vanishes at $V=V_{\rm c}$. 

%%%%%%%%%%%%%%%%%%%%%%%%%%%%%%%%%%%
\begin{figure}[tb]
\includegraphics[scale=0.55]{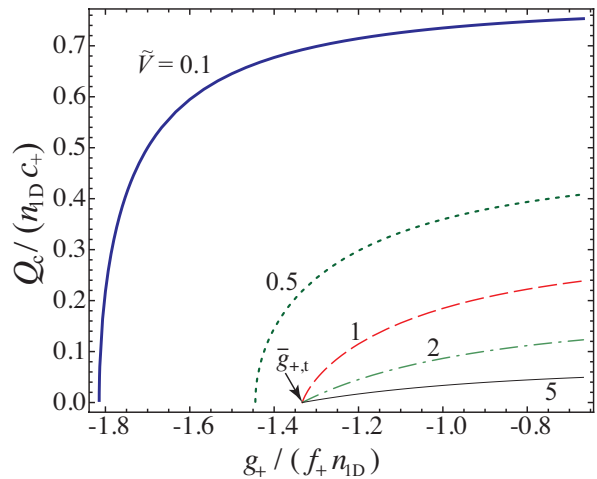}
\caption{\label{fig:qcg}
(color online) Critical current $Q_{\rm c}$ as a function of the barrier strength $V$ for several values of $\bar{g}_{+}$ across $\bar{g}_{+}=\bar{g}_{+,{\rm t}}$.
}
\end{figure}
%%%%%%%%%%%%%%%%%%%%%%%%%%%%%%%%%%%%

\begin{figure}[tb]
\includegraphics[scale=0.55]{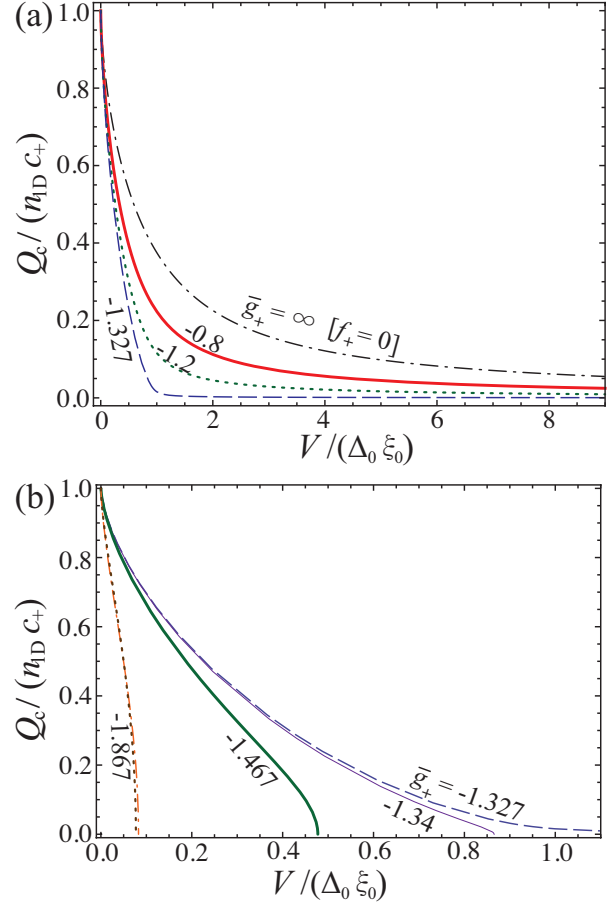}
\caption{\label{fig:qcv}
(color online) Critical current $Q_{\rm c}$ as a function of $\bar{g}_{+}$ for several values of $\tilde{V}$.
In (b), the dotted line represents the approximated value of Eq.~(\ref{eq:qc_ms}).
}
\end{figure}

For given values of $\bar{g}_{+}$ and $\tilde{V}$, there is the critical current $Q_{\rm c}$, above which a stable SF state is absent. The critical currents of a moving SF through a barrier potential have been measured in experiments~\cite{onofrio-00,engels-07, levy-07, ramanathan-11, desbuquois-12}. When $\bar{g}_{+}>\bar{g}_{+,{\rm t}}$ and $\tilde{V}\gg 1$, one can derive the critical current from Eq.~(\ref{eq:joseph}),
\begin{eqnarray}
\tilde{Q}_{\rm c} \simeq \frac{\gamma}{2(1+\gamma)\tilde{V}}.
\label{eq:qc_hv}
\end{eqnarray}
At the limit of $\bar{g}_{+} \rightarrow \infty$, the critical current is given by $\tilde{Q}_{\rm c} \simeq 1/(2\tilde{V})$, which agrees with the previous result for the GP equation~\cite{hakim-97, danshita-06}.
Equation (\ref{eq:qc_hv}) implies that the critical current vanishes as $\tilde{Q}_{\rm c} \propto \bar{g}_{+} - \bar{g}_{+,{\rm t}}\propto \frac{r_{\rm 1D}}{r_{\rm 1D, t}} - 1$ when the metastability limit at $\tilde{V}\ge1$  ($\bar{g}_{+}= \bar{g}_{+,{\rm t}}$) is approached. In Fig.~\ref{fig:qcg}, we plot $\tilde{Q}_{\rm c}$ as a function of $\bar{g}_{+}$ for several values of $\tilde{V}$. The dashed ($\tilde{V}=1$), dash-dotted ($\tilde{V}=2$), and thin-solid ($\tilde{V}=5$) lines indeed exhibit the linear dependence in a reasonably wide range near the transition point, in contrast to the thermodynamic quantities. This property is useful for identifying the criticality in experiments. 
In Fig.~\ref{fig:qcv}(a), we plot the critical current as a function of the barrier strength. When $\tilde{V}$ increases, $\tilde{Q}_{\rm c}$ monotonically decreases and its asymptotic behavior at $\tilde{V}\gg 1$ agrees well with Eq.~(\ref{eq:qc_hv}).  

In the case of the metastable SF state ($\bar{g}_{+,{\rm SF}}< \bar{g}_{+} < \bar{g}_{+,{\rm t}}$), assuming that $\bar{g}_{+}-\bar{g}_{+,\rm SF}\ll 1$, $\bar{q}\ll 1$, and $\bar{V}\ll 1$, one can analytically solve Eq.~(\ref{eq:z0det}) to obtain a useful expression,
\begin{eqnarray}
\tilde{Q}_{\rm c} \simeq 
\sqrt{1- \left(\frac{\sqrt{3}\tilde{V}}{\bar{g}_{+}-\bar{g}_{+,{\rm SF}}} \right)^{\frac{2}{3}}}.
\label{eq:qc_ms}
\end{eqnarray}
In the vicinity of the metastability limit, Eq.~(\ref{eq:qc_ms}) implies that $\tilde{Q}_{\rm c}\propto \sqrt{\bar{g}_{+} - \bar{g}_{+, {\rm c}}}$ for a fixed $V$ while $\tilde{Q}_{\rm c}\propto \sqrt{\tilde{V}_{\rm c} - \tilde{V}}$ for a fixed $\bar{g}_{+}$, where $\bar{g}_{+, {\rm c}}$ and $\tilde{V}_{\rm c}$ denote the values of $\bar{g}_{+}$ and $\tilde{V}$ at the metastability limit. In Fig.~\ref{fig:qcg}, the square-root dependence of the critical current is corroborated by the thick-solid ($\tilde{V}=0.1$) and dotted ($\tilde{V}=0.5$) lines near $\bar{g}_{+} = \bar{g}_{+,{\rm c}}$. In Fig.~\ref{fig:qcv}(b), $\tilde{Q}_{\rm c}$ versus $\tilde{V}$ is plotted and we see that the expression of Eq.~(\ref{eq:qc_ms}) represented by the dotted line well agrees with the precise numerical solution (the dash-dotted line). All the data at $\bar{g}_{+,{\rm SF}}<\bar{g}_{+}<\bar{g}_{+,{\rm t}}$ exhibit the square-root dependence near $V=V_{\rm c}$. Thus, in terms of the critical current, the barrier-induced criticality at $\tilde{V}\ge 1$ that is linear with respect to $\bar{g}_{+}-\bar{g}_{+,{\rm t}}$ is clearly distinguishable from the criticality at $\tilde{V}<1$.

%%%%%%%%%%%%%%%%%%%%%%%%%%%%%%%%%%%
%%%%%%%%%%%%%%%%%%%%%%%%%%%%%%%%%%%%

%%%%%%%%%%%%%%
\section{Conclusions}\label{sec:conc}
%%%%%%%%%%%%%%
We have studied superfluid (SF) Bose-Bose mixtures in optical lattices assuming that the hopping energy, the intra-component interaction, and the chemical potential for the two species are equal. On the basis of the sixth-order Ginzburg-Landau (GL) theory, we have shown that the SF state near the first-order quantum phase transition to the Mott insulator (MI) is described by the two-component nonlinear Schr\"dinger equation (NLSE) with cubic and quintic nonlinearity. We analyzed dark solitary-wave solutions of the cubic-quintic NLSE with a uniform potential to show that some properties of the solitary waves, such as the size and the inertial mass, exhibit critical behaviors near the first-order SF-MI transition. This criticality may be identified in experiments by measuring the velocity-phase relation.

Using the dark solitary-wave solutions, we have obtained the solution of the cubic-quintic NLSE with a barrier potential of $\delta$-function form. For the SF state that was metastable without the barrier, we have found  critical barrier strength above which the SF state is destabilized towards disjunction. 
Moreover, we discussed criticality near the new metastability limit induced by the strong barrier. We have derived the critical behavior of the critical current, which may be measured in experiments.
We also obtained the current-phase relation, and in particular we found its peculiar behavior for metastable SF states.

While we have focused on the dark solitary waves and barrier-potential effects in this paper, the cubic-quintic NLSE may be used to analyze other diverse effects and phenomena, including the formation of BEC droplets~\cite{gammal-00, enqvist-03} and nonlinear excitations in higher dimensions~\cite{barashenkov-88, barashenkov-89, malomed-05}. In this sense, the present work has opened up new possibilities for the studies of exotic nonlinear effects and phenomena in optical-lattice systems. 

%%%%%%%%%%%%%%%%%%%%%%%%%%%%%%%%%%%%%%%%%%%%%
%%%%%%%%%%%%%%%%%%%%%%%%%%%%%%%%%%%%%%%%%%%%%
\begin{acknowledgments}
%%%%%%%%%%%%%%%%%%%%%%%%%%%%%%%%%%%%%%%%%%%%%
The authors thank M.~Kunimi, G.~Marmorini, and G.~Watanabe for discussions.
The authors also thank the Yukawa Institute for Theoretical Physics (YITP) at Kyoto University for hospitality. Discussions during the YITP workshop (YITP-W-14-02) on ``Higgs Modes in Condensed Matter and Quantum Gases" were useful to complete this work. The authors acknowledge KAKENHI Grants from JSPS No.~25800228 (I.~D.), No.~25220711 (I.D.), No.~26800200 (D.Y.), and No.~26800199 (Y.K.).
\end{acknowledgments}

%%%%%%%%%%%%%%%%%


\begin{thebibliography}{99}
\bibitem{anderson-95}
M. H. Anderson, J. R. Ensher, M. R. Matthews, C. E. Wieman, and E. A. Cornell, 
Science {\bf 269}, 198 (1995).

\bibitem{davis-95}
K. B. Davis, M.-O. Mewes, M. R. Andrews, N. J. van Druten, D. S. Durfee, D. M. Kurn, and W. Ketterle, 
Phys. Rev. Lett. {\bf 75}, 3969 (1995).

\bibitem{dalfovo-99}
F. S. Dalfovo, L. P. Pitaevskii, S. Stringari, and S. Giorgini,
Rev. Mod. Phys. {\bf 71}, 463 (1999).

\bibitem{pitaevskii-03}
L. P. Pitaevskii and S. Stringari, 
{\it Bose-Einstein Condensation}
(Oxford University Press, Oxford, 2003).

\bibitem{steinhauer-02}
J. Steinhauer, R. Ozeri, N. Katz, and N. Davidson,
Phys. Rev. Lett. {\bf 88}, 120407 (2002).

\bibitem{hechenblaikner-00}
G. Hechenblaikner, O. M. Marag\`o, E. Hodby, J. Arlt, S. Hopkins, and C. J. Foot,
Phys. Rev. Lett. {\bf 85}, 692 (2000).

\bibitem{khaykovich-02}
L. Khaykovich, F. Schreck, G. Ferrari, T. Bourdel, J. Cubizolles, L. D. Carr, Y. Castin,  and C. Salomon,
Science {\bf 296}, 1290 (2002).

\bibitem{burger-99}
S. Burger, K. Bongs, S. Dettmer, W. Ertmer, K. Sengstock, A. Sanpera, G. V. Shlyapnikov, and M. Lewenstein,
Phys. Rev. Lett. {\bf 83}, 5198 (1999).

\bibitem{denschlag-00}
J. Denschlag, J. E. Simsarian, D. L. Feder, C. W. Clark, L. A. Collins, J. Cubizolles, L. Deng, E. W. Hagley, K. Helmerson, W. P. Reinhardt, S. L. Rolston, B. I. Schneider, and W. D. Phillips,
Science {\bf 287}, 97 (2000).

\bibitem{becker-08}
C. Becker, S. Stellmer, P. Soltan-Panahi, S. D\"orscher, M. Baumert, E.-M. Richter, J. Kronj\"ager, K. Bongs, and K. Sengstock,
Nat. Phys. {\bf 4}, 496 (2008).

\bibitem{madison-01}
K. W. Madison, F. Chevy, V. Bretin, and J. Dalibard,
Phys. Rev. Lett. {\bf 86}, 4443 (2001).

\bibitem{abo-02}
J. R. Abo-Shaeer, C. Raman, J. M. Vogels, and W. Ketterle,
Science {\bf 292}, 476 (2002).

\bibitem{onofrio-00}
R. Onofrio, C. Raman, J. M. Vogels, J. R. Abo-Shaeer, A. P. Chikkatur, and W. Ketterle,
Phys. Rev. Lett. {\bf 85}, 2228 (2000).

\bibitem{sarlo-05}
L. De Sarlo, L. Fallani, J. E. Lye, M. Modugno, R. Saers, C. Fort, and M. Inguscio, 
Phys. Rev. A {\bf 72}, 013603 (2005).

\bibitem{engels-07}
P. Engels and C. Atherton,
Phys. Rev. Lett. {\bf 99}, 160405 (2007).

\bibitem{levy-07}
S. Levy, E. Lahoud, I. Shomroni, and J. Steinhauer,
Nature (London) {\bf 449}, 579 (2007).

\bibitem{ramanathan-11}
A. Ramanathan, K. C. Wright, S. R. Muniz, M. Zelan, W. T. Hill, C. J. Lobb, K. Helmerson, W. D. Phillips, and G. K. Campbell,
Phys. Rev. Lett. {\bf 106}, 130401 (2011).

\bibitem{desbuquois-12}
R. Desbuquois, L. Chomaz, T. Yefsah, J. L\'eonard, J. Beugnon, C. Weitenberg, and J. Dalibard,
Nat. Phys. {\bf 8}, 645 (2012).

\bibitem{albiez-05}
M. Albiez, R. Gati, J. F\"olling, S. Hunsmann, M. Cristiani, and M. K. Oberthaler, 
Phys. Rev. Lett. {\bf 95}, 010402 (2005).

\bibitem{greiner-02}
M. Greiner, O. Mandel, T. Esslinger, T. W. H\"ansch, and I. Bloch, 
Nature (London) {\bf 415}, 39 (2002).

\bibitem{fisher-89}
M. P. A. Fisher, P. B. Weichman, G. Grinstein, and D. S. Fisher, 
Phys. Rev. B {\bf 40}, 546 (1989).

\bibitem{altman-02}
E. Altman and A. Auerbach,
Phys. Rev. Lett. {\bf 89}, 250404 (2002).

\bibitem{endres-12}
M. Endres, T. Fukuhara, D. Pekker, M. Cheneau, P. Schau\ss, C. Gross, E. Demler, S. Kuhr, and I. Bloch,
Nature (London) {\bf 487}, 454 (2012). 

\bibitem{pekker-14}
D. Pekker and C. M. Varma,
arXiv:1406.2968 (2014).

\bibitem{altman-05}
E. Altman, A. Polkovnikov, E. Demler, B. I. Halperin, and M. D. Lukin,
Phys. Rev. Lett. {\bf 95}, 020402 (2005).

\bibitem{mun-07}
J. Mun, P. Medley, G. K. Campbell, L. G. Marcassa, D. E. Pritchard, and W. Ketterle, 
Phys. Rev. Lett. {\bf 99}, 150604 (2007).

\bibitem{paredes-04}
B. Paredes, A. Widera, V. Murg, O. Mandel, S. F\"olling, I. Cirac, G. V. Shlyapnikov, T. W. H\"ansch, and I. Bloch,
Nature (London) {\bf 429}, 277 (2004).

\bibitem{balakrishnan-09}
R. Balakrishnan, I. I. Satija, and C. W. Clark,
Phys. Rev. Lett. {\bf 103}, 230403 (2009).

\bibitem{kato-14}
Y. Kato, D. Yamamoto, and I. Danshita,
Phys. Rev. Lett. {\bf 112}, 055301 (2014).

\bibitem{ginzburg-82}
V. L. Ginzburg and A. A. Sobyanin,
J. Low Temp. Phys. {\bf 49}, 507 (1982).

\bibitem{lipowsky-82}
R. Lipowsky,
Phys. Rev. Lett. {\bf 49}, 1575 (1982).

\bibitem{lipowsky-83}
R. Lipowsky and W. Speth,
Phys. Rev. B {\bf 28}, 3983 (1983).

\bibitem{sornette-85}
D. Sornette,
Phys. Rev. B {\bf 31}, 4672 (1985).

\bibitem{binder-87}
K. Binder,
Rep. Prog. Phys. {\bf 50} 783 (1987).

\bibitem{barashenkov-88}
I. V. Barashenkov and V. G. Makhankov,
Phys. Lett. A {\bf 128}, 52 (1988). 

\bibitem{barashenkov-89}
I. V. Barashenkov, A. D. Gocheva, V. G. Makhankov, and I. V. Puzynin,
Physica D {\bf 34}, 240 (1989).

\bibitem{gagnon-89}
L. Gagnon,
J. Opt. Soc. Am. A {\bf 6}, 1477 (1989).

\bibitem{kivshar-98}
Y. S. Kivshar and B. Luther-Daviews,
Phys. Rep. {\bf 298}, 81 (1998).

\bibitem{malomed-05}
B. A. Malomed, D. Mihalache, F. Wise, and L. Torner,
J. Opt. B: Quantum Semiclass. Opt. {\bf 7}, R53 (2005).

\bibitem{johnson-09}
P. R. Johnson, E. Tiesinga, J. V. Porto, and C. J. Williams,
New J. Phys. {\bf 11}, 093022 (2009).

\bibitem{mahmud-13}
K. W. Mahmud and E. Tiesinga,
Phys. Rev. A {\bf 88}, 023602 (2013).

\bibitem{petrov-14}
D. S. Petrov,
Phys. Rev. Lett. {\bf 112}, 103201 (2014).

\bibitem{frantzeskakis-10}
D. J. Frantzeskakis,
J. Phys. A {\bf 43}, 213001 (2010).

\bibitem{scott-11}
R. G. Scott, F. Dalfovo, L. P. Pitaevskii, and S. Stringari,
Phys. Rev. Lett. {\bf 106}, 185301 (2011).

\bibitem{pitaevskii-14}
L. P. Pitaevskii,
arXiv:1407.8081 (2014).

\bibitem{yefsah-13}
T. Yefsah, A. T. Sommer, M. J. H. Ku, L. W. Cheuk, W. Ji, W. S. Bakr, and M. W. Zwierlein,
Nature (London) {\bf 499}, 426 (2013).

\bibitem{ku-14}
M. J. H. Ku, W. Ji, B. Mukherjee, E. Guardado-Sanchez, L. W. Cheuk, T. Yefsah, and M. W. Zwierlein,
Phys. Rev. Lett. {\bf 113}, 065301 (2014).

\bibitem{lipowsky-84}
R. Lipowsky and G. Gompper,
Phys. Rev. B {\bf 29}, 5213 (1984).

\bibitem{lipowsky-87}
R. Lipowsky,
Ferroelectrics {\bf 73}, 69 (1987).

\bibitem{josephson-62}
B. D. Josephson,
Phys. Lett. {\bf 1}, 251 (1962).

\bibitem{baratoff-70}
A. Baratoff, J. A. Blackburn, and B. B. Schwartz, 
Phys. Rev. Lett. {\bf 25}, 1096 (1970).

\bibitem{danshita-06}
I. Danshita, N. Yokoshi, and S. Kurihara,
New J. Phys. {\bf 8}, 44 (2006).

\bibitem{watanabe-09}
G. Watanabe, F. Dalfovo, F. Piazza, L. P. Pitaevskii, and S. Stringari,
Phys. Rev. A {\bf 80}, 053602 (2009).

\bibitem{jaksch-98}
D. Jaksch, C. Bruder, J. I. Cirac, C. W. Gardiner, and P. Zoller,
Phys. Rev. Lett. {\bf 81}, 3108 (1998).

\bibitem{hall-98}
D. S. Hall, M. R. Matthews, J. R. Ensher, C. E. Wieman, and E. A. Cornell,
Phys. Rev. Lett. {\bf 81}, 1539 (1998).

\bibitem{egorov-13}
M. Egorov, B. Opanchuk, P. Drummond, B. V. Hall, P. Hannaford, and A. I. Sidorov,
Phys. Rev. A {\bf 87}, 053614 (2013).

\bibitem{widera-08}
A. Widera, S. Trotzky, P. Cheinet, S. F\"olling, F. Gerbier, I. Bloch, V. Gritsev, M. D. Lukin, and E. Demler,
Phys. Rev. Lett. {\bf 100}, 140401 (2008).

\bibitem{tojo-10}
S. Tojo, Y. Taguchi, Y. Masuyama, T. Hayashi, H. Saito, and T. Hirano, 
Phys. Rev. A {\bf 82}, 033609 (2010).

\bibitem{mckay-10}
D. McKay and B. DeMarco, 
New J. Phys. {\bf 12}, 055013 (2010).

\bibitem{kuklov-03}
A. B. Kuklov and B. V. Svistunov,
Phys. Rev. Lett. {\bf 90}, 100401 (2003).

\bibitem{paredes-03}
B. Paredes and J. I. Cirac, 
Phys. Rev. Lett. {\bf 90}, 150402 (2003).

\bibitem{chen-03}
G.-H. Chen and Y.-S. Wu, 
Phys. Rev. A {\bf 67}, 013606 (2003).

\bibitem{altman-03}
E. Altman, W. Hofstetter, E. Demler, and M. D. Lukin, 
New J. Phys. {\bf 5}, 113 (2003).

\bibitem{kuklov-04-a}
A. Kuklov, N. Prokof'ev, and B. Svistunov, 
Phys. Rev. Lett. {\bf 92}, 050402 (2004).

\bibitem{kuklov-04-b}
A. Kuklov, N. Prokof'ev, and B. Svistunov, 
Phys. Rev. Lett. {\bf 92}, 030403 (2004).

\bibitem{isacsson-05}
A. Isacsson, M.-C. Cha, K. Sengupta, and S. M. Girvin, 
Phys. Rev. B {\bf 72}, 184507 (2005).

\bibitem{arguelles-07}
A. Arg\"uelles and L. Santos, 
Phys. Rev. A {\bf 75}, 053613 (2007).

\bibitem{mishra-07}
T. Mishra, R. V. Pai, and B. P. Das, 
Phys. Rev. A {\bf 76}, 013604 (2007).

\bibitem{mathey-09}
L. Mathey, I. Danshita, and C. W. Clark, 
Phys. Rev. A {\bf 79}, 011602(R) (2009).

\bibitem{hu-09}
A. Hu, L. Mathey, I. Danshita, E. Tiesinga, C. J. Williams, and C. W. Clark, 
Phys. Rev. A {\bf 80}, 023619 (2009).

\bibitem{hubener-09}
A. Hubener, M. Snoek, and W. Hofstetter, 
Phys. Rev. B 80, 245109 (2009).

\bibitem{iskin-10}
M. Iskin, 
Phys. Rev. A 82, 033630 (2010).

\bibitem{chen-10}
P. Chen and M. F. Yang, 
Phys. Rev. B 82, 180510(R) (2010).

\bibitem{sansone-10}
B Capogrosso-Sansone, \c{S}. G. S\"oyler, N. V Prokof’ev, and B. V Svistunov,
Phys. Rev. A {\bf 81}, 053622 (2010).

\bibitem{ozaki-12}
T. Ozaki, I. Danshita, and T. Nikuni,
arXiv:1210.1370.

\bibitem{li-13}
Y. Li, L. He, and W. Hofstetter,
New J. Phys. {\bf 15}, 093028 (2013).

\bibitem{yamamoto-13}
D. Yamamoto, T. Ozaki, C. A. R. S\'a de Melo, and I. Danshita, 
Phys. Rev. A {\bf 88}, 033624 (2013).

\bibitem{folling-06}
S. F\"olling, A. Widera, T. M\"uller, F. Gerbier, and I. Bloch, 
Phys. Rev. Lett. {\bf 97}, 060403 (2006).

\bibitem{campbell-06}
G. K. Campbell, J. Mun, M. Boyd, P. Medley, A. E. Leanhardt, L. G. Marcassa, D. E. Pritchard, and W. Ketterle, 
Science {\bf 313}, 649 (2006).

\bibitem{gemelke-09}
N. Gemelke, X. Zhang, C.-L. Hung, and C. Chin,
Nature (London) {\bf 460}, 995 (2009).

\bibitem{schurmann-00}
H. W. Sch\"urmann and V. S. Serov,
Phys. Rev. E {\bf 62}, 2821 (2000).

\bibitem{tsuzuki-71}
T. Tsuzuki, 
J. Low Temp. Phys. {\bf 4}, 441 (1971).

\bibitem{hakim-97}
V. Hakim,
Phys. Rev. E {\bf 55}, 2835 (1997).

\bibitem{gammal-00}
A. Gammal, T. Frederico, L. Tomio, and Ph. Chomaz,
J. Phys. B {\bf 33}, 4053 (2000).

\bibitem{enqvist-03}
K. Enqvist and M. Laine,
J. Cosmol. Astropart. Phys. {\bf 08}, 003 (2003).

\end{thebibliography}
\end{document}